\documentclass[
,twocolumn
]{aastex631}

\usepackage{amsmath, subfigure}
\usepackage{threeparttable}
\usepackage{orcidlink}
\usepackage{url}

\providecommand{\DIFdel}[1]{}%

\begin{document}

\title{Forecasting Supermassive  Black Hole Binary Gravitational Wave Probes: Prospects for Future Pulsar Timing Array and Space-Borne Detectors}

\author{Katsunori Kusakabe}
\affiliation{Department of Earth and Space Science, Graduate School of Science, The University of Osaka, 1-1 Machikaneyama, Toyonaka, Osaka 560-0043, Japan}

\author[0000-0002-7272-1136]{Yoshiyuki Inoue}    
\affiliation{Department of Earth and Space Science, Graduate School of Science, The University of Osaka, 1-1 Machikaneyama, Toyonaka, Osaka 560-0043, Japan}
\affiliation{Interdisciplinary Theoretical \& Mathematical Science Center (iTHEMS), RIKEN, 2-1 Hirosawa, 351-0198, Japan}
\affiliation{Kavli Institute for the Physics and Mathematics of the Universe (WPI), UTIAS, The University of Tokyo, 5-1-5 Kashiwanoha, Kashiwa, Chiba 277-8583, Japan}

\author[0000-0003-3467-6079]{Daisuke Toyouchi}
\affiliation{Department of Earth and Space Science, Graduate School of Science, The University of Osaka, 1-1 Machikaneyama, Toyonaka, Osaka 560-0043, Japan}

\author[0000-0002-3034-5769]{Keitaro Takahashi}
\affiliation{Faculty of Advanced Science and Technology, Kumamoto University, Kumamoto 860-8555, Japan}
\affiliation{International Research Organization for Advanced Science and Technology, Kumamoto University, Kumamoto 860-8555, Japan}

\correspondingauthor{Katsunori Kusakabe}
\email{u778278e@ecs.osaka-u.ac.jp}

\begin{abstract}

We present a comprehensive framework for predicting the detection prospects of supermassive black hole binaries (SMBHBs) by future gravitational wave (GW) observatories, examining both space-borne detectors (LISA, Taiji, TianQin) and next-generation pulsar timing array (PTA) combined with the Square Kilometre Array (SKA-PTA). Leveraging dual active galactic nucleus (AGN) fractions and AGN X-ray luminosity functions, we systematically evaluate the detectable SMBHB populations with a detection threshold of signal-to-noise ratio $\geq 5$ for each GW observatory. Our analysis reveals that space-borne detectors are expected to identify approximately $\sim 1\text{--}2$  to $\sim$ 20 events per year, depending on the SMBHB orbital evolution prescriptions. On the other hand, SKA-PTA demonstrates the potential to reach the first GW detection from individual SMBHBs within a few years of observation and achieve detectable GW source counts of $10^2 \text{--} 10^3$ after about 10 years, depending on PTA configurations. These facilities will significantly improve SMBHB detectability and enable characterization of their properties across different frequency bands.

\end{abstract}


\section{\label{sec:intro}INTRODUCTION}
The formation and evolution of supermassive black hole binaries (SMBHBs) play a fundamental role in our understanding of galaxy evolution, structure formation, and cosmic history. These systems, formed through galaxy mergers, represent a crucial phase in the growth of supermassive black holes (SMBHs). While direct electromagnetic (EM) observations of these systems at sub-pc scales are challenging due to extraordinary resolution requirements \citep{Burke-Spolaor:2018bvk, DOrazio:2023rvl}, the new observational windows by gravitational waves (GWs) -- namely Pulsar Timing Array (PTA) \citep{1990ApJ...361..300F} or space-borne interferometers such as LISA \citep{LISA:2022yao}, TianQin \citep{TianQin:2015yph}, and Taiji \citep{Chen:2023zkb} -- are expected to explore SMBHB characterization at different GW frequency regimes. 

In the nHz band, PTA collaborations \citep{NANOGrav:2023hde, NANOGrav:2023gor, EPTA:2023sfo, EPTA:2023xxk, Zic:2023gta, Reardon:2023gzh, Xu:2023wog} have opened unprecedented observational windows into the realm of SMBHB systems through their detection of evidence for the stochastic gravitational wave background (SGWB). This evidence marks the first indication of low-frequency gravitational waves, suggesting a substantial population of evolving SMBHBs throughout cosmic history. While targeted searches in EPTA and PPTA data have found no significant detections \citep{EPTA:2023gyr, Zhao:2025pgg}, recent analyses based on the NANOGrav 15 yr dataset have identified two marginal SMBHB candidates \citep{Agarwal:2025cag}, suggesting we are approaching the threshold of individual source detection. The forthcoming PTAs, particularly those integrated with the Square Kilometre Array (hereafter SKA-PTA), promise a quantum leap in sensitivity \citep{Smits:2008cf, Janssen:2014dka}. By expanding the millisecond pulsar (MSP) network and significantly improving timing precision, this next-generation array is expected to detect a substantial population of GW emissions from individual GW sources \citep{Feng:2020nyw, Chen:2023hyj}.

In the mHz band, space-borne interferometers are expected to enable the detection of SMBHBs across a wide range of cosmic redshifts and SMBH masses ($\mathcal{O}(10^3\text{-}10^9\ M_{\odot}$)) \citep{Amaro-Seoane:2012vvq, Burke-Spolaor:2018bvk, Gong:2021gvw, LISA:2024hlh}, providing crucial constraints on SMBH binary demographics, merger rates, and binary formation efficiency. Observations of the final inspiral and merger phases will enable precise measurements of binary parameters (masses, mass ratios, spins, luminosity distances) that are challenging with EM observations alone \citep{Burke-Spolaor:2018bvk, LISA:2024hlh}. High signal-to-noise ratio (S/N) detections of nearby merging events will also enable tests of general relativity in the strong-field regime \citep{Yunes:2013dva, Berti:2015itd} and cosmological distance measurements through GW standard sirens \citep{Holz:2005df, LISACosmologyWorkingGroup:2019mwx}. These detectors complement ongoing PTA campaigns, enabling multi-band GW astronomy and potential co-detection of systems at different evolutionary stages \citep{Sesana:2008mz, Spallicci:2011nr, Burke-Spolaor:2018bvk, Ellis:2023owy}.

The GW signals detected by these facilities will provide crucial insights into evolutionary pathways of SMBHs through mergers. Previous studies have investigated the detectability of resolvable GWs from individual SMBHB systems  \citep{Sesana:2010wy, Klein:2015hvg, Salcido:2016oor, Bonetti:2018tpf, Katz:2019qlu, 2019MNRAS.486.2336D, Barausse:2020mdt, Feng:2020nyw, Curylo:2021pvf, Chen:2023hyj, 2024A&A...686A.183I}, typically relying on SMBHB merger rates inferred from galaxy or dark matter halo mergers through cosmological simulations \citep{Salcido:2016oor, Katz:2019qlu} or semi-analytical models (SAMs) based on galaxy mergers \citep{Sesana:2010wy, Klein:2015hvg, Bonetti:2018tpf, 2019MNRAS.486.2336D, Barausse:2020mdt, Feng:2020nyw, Curylo:2021pvf, Chen:2023hyj, 2024A&A...686A.183I}. However, these predictions exhibit significant variations depending on the assumed merger timescales, environmental effects, and binary formation efficiencies (see \citet{2019astro2020T.504K} and references therein). 

To address the construction of SMBHB population synthesis, recent studies have proposed alternative approaches based on quasar observations~\citep{2009ApJ...700.1952H, Goulding:2019hnn, Xin:2021mmk, Casey-Clyde:2021xro, Kis-Toth:2024gkm, Xin:2025voy, Lapi:2025wxt}. While these observationally motivated models provide a more direct connection to SMBHB populations, they face challenges in determining quasar pair formation rates. More straightforward constraints come from recent breakthroughs in dual active galactic nucleus (AGN) studies~\citep{2011ApJ...737..101L, 2012ApJ...746L..22K, 2020ApJ...899..154S, Shen:2022cmp, 2023arXiv231003067P, 2024arXiv240514980L}, which have provided direct estimations on SMBHB occurrence rates ranging from $\sim0.01$ \% to $\sim20$ \%, enabling us to probe SMBHB populations directly. 

In our previous work~\citep{Kusakabe:2025rmn}, we developed the refined SMBHB population model that incorporates these observed dual AGN fractions with the AGN X-ray luminosity function~\citep{Ueda:2014tma}. The model reproduced the recently detected nHz SGWB spectrum, deciphering the relationship between GW and EM observations. Building on this framework, in this paper, we investigate the detectability of individual SMBHB systems by future GW observatories through a comprehensive S/N analysis.

We present our methodology for estimating detectable SMBHB populations, including S/N analysis and detection threshold criteria in Section \ref{sec:method}. Section \ref{sec:result} indicates our findings of the SMBHB detection rates for both space-based detectors and SKA-PTA. In Section \ref{sec:discussion}, we discuss the comparison with previous works, uncertainties of the detection rates, and implications for the future GW probes, and conclusions are in Section \ref{sec:conclusion}. Throughout this work, we adopt standard cosmological parameters with $(H_0, \Omega_m, \Omega_{\Lambda})= (67.4~ \mathrm{km}~ \mathrm{s}^{-1}~ \mathrm{Mpc}^{-1}, 0.315, 0.685)$ \citep{Planck:2018vyg}.

\section{\label{sec:method}Estimation on the Detectable SMBHB population}
\subsection{\label{sec:S/N} Formalization of Detectable GW Sources}
SMBHBs emit GWs across different frequency bands throughout their evolutionary stages (inspiral-merger-ringdown). To evaluate the number of detectable sources, we employ the SMBHB merger rate ${d^3 n}/{dMdzdq}$ derived from our previous population model in \citet{Kusakabe:2025rmn}. Here $z$ denotes redshift, $M$ represents the primary SMBH mass, $q$ is the mass ratio, and $n$ is the comoving number density of sources. We first represent the coalescence rate per year $dn/dt$ in terms of the differential source count per logarithmic frequency and binary parameters $d^4 n/dMdzdqd\mathrm{log}f$ written as \citep{Sesana:2008mz, Chen:2023hyj, Sato-Polito:2024lew}

\begin{equation}
\frac{d^4 n}{dMdzdqd\mathrm{log}f} =  \frac{d^3 n}{dMdzdq} \frac{dt_r}{d\mathrm{log}f_r} \frac{dz}{dt_r} \frac{dV_c}{dz} 
\label{eq:gw_total}
\end{equation}
where 
\begin{equation}
\frac{dz}{dt_r} \frac{dV_c}{dz} = \frac{4\pi cd_{\mathrm{L}}^2(z)}{(1+z)}.    
\end{equation}
Here, $V_c$ denotes the comoving volume, while $t$ and $t_r$ correspond to the coordinate time in the observer's and source's rest frames, respectively. The luminosity distance between the source and the observer is represented by $d_{\mathrm{L}}$. The GW frequency determined by the observer is $f$, while the source frequency is expressed as $f_r = (1 + z)f$. We assume the binary orbit is circular throughout our analysis. The frequency evolution term $d\mathrm{log}f_r / d t_r$ for a binary evolving through GW radiation is then given by  \citep{Sesana:2008mz, Chen:2023hyj, Sato-Polito:2024lew}
\begin{equation}
\frac{d\mathrm{log}f_r}{d t_r} = \frac{96\pi^{8/3}}{5} \Big(\frac{GM}{c^3}\Big)^{5/3} \eta f^{8/3}_{r}
\end{equation}
where the symmetric mass ratio is defined as $\eta \equiv q/(1+q)^2$. Our analysis spans the parameter space $0\leq z \leq 5$, $10^5 M_{\odot} \leq M \leq 10^{11} M_{\odot}$, and $0.1 \leq q \leq 1$, consistent with \citet{Kusakabe:2025rmn}. As for S/N, we adopt the sky location, inclination, and polarization averaged S/N: $\rho$ described in detail in Section \ref{sec:threshold} and implement a detection threshold of $\rho_{\mathrm{crit}} = 5$ through the step function $\theta(\rho- \rho_{\mathrm{crit}})$, which selects detectable sources satisfying $\rho \geq \rho_{\mathrm{crit}}$. 

In case of the PTA detection rate, the relevant frequency range is $f \in [{T_{\mathrm{obs}}}^{-1},\Delta t^{-1}]$ Hz where $T_{\rm{obs}}$ is the observational period of the PTA and $1 / \Delta t$ is the observational cadence (see Section~\ref{sec:threshold} for the discrete parameter settings). As discussed in \citet{Chen:2023hyj}, we assume that the comoving merger rate density $d^3n/dzdMdq$ remains constant during the SMBHB frequency evolution from emitting GW at the frequency of $f_r$ to its final coalescence. The detection rate of GW sources per year for PTAs is then obtained by integrating Eq.~\ref{eq:gw_total} with imposing a step function as

\begin{equation}
\frac{dn}{dt} = \int dzdMdqd\mathrm{log}f \frac{d^4 n}{dMdzdqd\mathrm{log}f} \theta(\rho - \rho_{\mathrm{crit}})\ [\rm{yr}^{-1}].
\label{eq:pta_detection}
\end{equation}

In contrast, space-borne detectors are designed to observe GWs during the final stages of binary evolution, from the late inspiral through merger, typically within years to days before coalescence. For these short-lived events, the observable event rate is directly determined by the intrinsic merger rate $d^3n/dMdzdq$ expressed as \citep{Arun:2008zn, Curylo:2021pvf, Furusawa:2023fwl}

\begin{equation}
\frac{dn}{dt} = \int dzdMdq \frac{d^3 n}{dMdzdq} \frac{dz}{dt_r} \frac{dV_{c}}{dz} \theta(\rho  - \rho_{\mathrm{crit}})\ [\rm{yr}^{-1}] .
\label{eq:step}
\end{equation}

For space-borne detector predictions, we consider two scenarios for the initial orbital separation, which sets the starting point of the merger timescale calculation (see \citet{Kusakabe:2025rmn} for detailed merger timescale prescriptions). The merger timescale is predominantly governed by dynamical friction during the early stages of orbital evolution, making the choice of initial separation a key determinant of the predicted detection rates. For massive SMBHBs targeted by SKA-PTA (typically $\gtrsim 10^{8-9}\,M_{\odot}$), dynamical friction timescales are relatively short compared to lighter systems, rendering the choice of initial separation less influential on detection predictions. Accordingly, we adopt a fixed typical initial separation of $a_0 = 10$ kpc for PTA calculations. However, for the lower-mass systems ($\lesssim 10^7\,M_{\odot}$) accessible to space-borne detectors, dynamical friction timescales become more sensitive to the initial separation, potentially affecting merger rate predictions. We therefore consider two scenarios for space-borne detectors. In the first scenario, we adopt a fixed effective radius of $a_0 = 10$ kpc for all host galaxies, corresponding to typical stellar masses of $M_{\star} \sim 10^{11} M_{\odot}$. This choice serves as a baseline consistent with our previous work~\citep{Kusakabe:2025rmn}. The other scenario incorporates a mass-dependent initial separation based on empirical scaling relations between galaxy effective radius and stellar mass. This approach recognizes that the beginning of the SMBH merger should scale with the characteristic size of the progenitor galaxies, which in turn correlates with their stellar masses. According to \citet{Lange:2014kca}, the relationship between the total stellar mass of a galaxy, $M_{\mathrm{gal}}$, and its effective half-light radius in kpc, $R_e$, is described by a double power law witten as
\begin{equation}
R_e = \gamma \left(\frac{M_{\mathrm{gal}}}{M_\odot}\right)^{\alpha}
      \left(1 + \frac{M_{\mathrm{gal}}}{M_0}\right)^{(\beta - \alpha)},
\end{equation}
where $(\alpha, \beta, \gamma, M_0) = (0.11, 0.76, 0.11, 2.01\times10^{10}\,M_\odot)$. Here, early‐type galaxies are chosen since they are likely to experience major mergers, therefore likely to be dominant GW sources. To convert this relation to SMBH mass, we use scaling laws between galaxy mass and bulge mass \citep{Sesana:2013wja,Kormendy:2013dxa}: 
\begin{align}
& \frac{M_{\rm{bulge}}}{M_{\rm{gal}}} = && \nonumber \\
& \quad
\begin{cases}
0.615 & \hspace{-3mm} (\log_{10} M_{\rm{gal}} < 10) \\
\frac{\sqrt{6.9}}{(\log_{10} M_{\rm{gal}} - 10)^{1.5}} \exp\left(-\frac{3.45}{\log_{10} M_{\rm{gal}} - 10}\right) \\  
+ 0.615 &  \hspace{-3mm} (\log_{10} M_{\rm{gal}} \ge 10)
\end{cases}
\end{align}
and the relationship between BH and bulge mass $M = 10^{-2.83} {M_{\rm{bulge}}}$ \citep{Kormendy:2013dxa}. Here, we take the fixed separation model of $a_0$ = 10 kpc and explore the mass-dependent case, referred to as the mass-dependent $a_0$ model. 

\subsection{\label{sec:threshold}Evaluation of Detection Threshold of GW Surveys}
This section briefly compiles the S/N analysis for each GW detector described in the literature (e.g., \citet{Robson:2018ifk, Kaiser:2020tlg}). We first introduce the general S/N formalism applicable to both PTAs and space-borne detectors. The averaged S/N is given by \citep{Robson:2018ifk}
\begin{equation} 
\rho^2 = \int_{f=0}^{\infty}  \frac{{4|\tilde{h}(f)|^2}}{{S_n(f)}} df.
\end{equation}
Here, $S_n(f)$ is the effective noise power spectral density (PSD), defined as $S_n(f) = P_n(f)/\mathcal{R}(f)$, where $P_n(f)$ is the PSD of the detector noise and $\mathcal{R}(f)$ is the instrument response function. The GW signal is characterized in the frequency domain by the complex strain amplitude $\tilde{h}(f)$, which is the Fourier transform of the time-domain strain $h(t)$. For binary BH coalescence, we model $\tilde{h}(f)$ using the phenomenological waveform model \citep{Khan:2015jqa, Husa:2015iqa}, which captures the inspiral, merger,  and ringdown phases in the frequency domain \citep{Moore:2014lga, Robson:2018ifk, Kaiser:2020tlg}. In the following, we describe the S/N calculations based on different noise sources for the space-borne detector and the PTA, respectively.

PTA sensitivity is influenced by multiple noise sources, including time-uncorrelated white noise from radiometer noise and pulse jitter \citep{2011ASSP...21..229H, Thrane:2013oya}, red noise from the pulsar spin variations \citep{Kaiser:2020tlg, EPTA:2023akd}, and interstellar medium effects such as dispersion measure (DM) variations \citep{EPTA:2023akd, Babak:2024yhu}, contributions from the stochastic gravitational wave background (SGWB) and other inhibiting components \citep{EPTA:2015ike, Babak:2024yhu}. Our analysis primarily focuses on the white noise contributions to the timing root-mean-square (RMS) as analogous to the previous works \citep{Feng:2020nyw, Chen:2023hyj}, while the impact of red noise and other inhibiting components are discussed in Section~\ref{subsec:red_noise}. For PTA configurations, the sky and polarization averaged response function is given by \citep{Hazboun:2019vhv}
\begin{equation}
\mathcal{R}(f) = \frac{1}{12\pi^2 f^2}.   
\end{equation}
The PSD of the PTA observations can be denoted as the sum of the PSD from each pulsar with the number of monitoring MSPs $N_{\mathrm{pl}}$ as \citep{Thrane:2013oya, Lam:2018uta}
\begin{equation}
P_n(f) = \sum_{i}^{N_{\mathrm{pl}}} P_{\mathrm{WN},i}(f)   
\label{eq:WNPSD}
\end{equation}
where the white noise contribution from each pulsar is
\begin{equation}
P_{\mathrm{WN},i} = 2\Delta t_i {\sigma_{t_i}}^2.     
\end{equation}
Here, $\sigma_{t_i}$ and $1 / \Delta t_i$ represent the RMS timing residual and the observational cadence of each pulsar. For simplicity, we assume uniform values $\sigma_{t_i} = \sigma_t$ and $\Delta t_i = \Delta t$ across all pulsars. These parameters, together with the number of pulsars $N_{\mathrm{pl}}$, determine the overall sensitivity of the PTA. To assess the capabilities of future pulsar timing facilities, we adopt parameters anticipated for the next-generation PTA. Based on projections in \citet{Porayko:2018sfa}, SKA-PTA are expected to achieve timing residuals below 50 ns for well-timed MSPs, with an average of around 30 ns. The highest-precision pulsars are projected to reach $\sigma_t \sim 10$ ns, significantly outperforming the typical timing accuracy of the order of 0.1--1~\textmu s currently achieved in PTA observations. For our fiducial SKA-PTA model, we adopt fiducial parameters ($\sigma_t, N_{\mathrm{pl}}, \Delta t$) = (30 ns, 500, 0.02 yr). We also explore parameter ranges of $\sigma_t =(10,30,50)$ ns, $N_{\mathrm{pl}} = (100, 500, 1000)$ for the timing precision and the number of pulsars used in PTA observations, with fixed observational cadence $\Delta t =0.02$ yr \citep{Chen:2023hyj, Guo:2024tlg}. For our SKA-PTA sensitivity analysis, we employ the \href{https://github.com/ark0015/gwent}{\texttt{gwent}} software package \citep{Kaiser:2020tlg}, which calculates signal-to-noise ratios across broadband frequencies for arbitrary SMBHB parameter configurations, including PTA measurements. The package implements the IMRPhenomD phenomenological waveform model \citep{Khan:2015jqa, Husa:2015iqa}, which incorporates BH spin parameters. Given the relatively weak spin dependence noted in \citet{Kaiser:2020tlg}, we set the spin parameter to zero throughout our computations.

On the other hand, space-based GW detectors encounter three primary noise sources: interferometer (IFO) noise, acceleration noise, and galactic confusion noise \citep{2017arXiv170200786A}. IFO noise stems from shot noise and optical path disturbances in system components, while acceleration noise results from perturbations affecting the test mass's free-fall motion in the spacecraft. Galactic confusion noise originates from unresolved binary systems in our Galaxy, which collectively complicate the detection of individual SMBHB signals. The effective PSD including these noise components is given by \citep{2017arXiv170200786A}

\begin{align}
S_{n}(f) &= \frac{10}{3\mathcal{R}(f)}\Big(\frac{P_{\mathrm{IFO}}} {L_{\mathrm{arm}}^{2}}+2(1+\cos^{2}\left(f/f_{\ast}\right)) \nonumber \\
&\hspace{30mm} \times \frac{P_{\mathrm{acc}}}{(2\pi f)^{4}L_{\mathrm{arm}}^{2}}\Big) + S_c(f)
\end{align}
where $P_{\mathrm{IFO}}$ is the PSD of IFO noise, $P_{\mathrm{acc}}$ is the PSD of acceleration noise, $L_{\mathrm{arm}}$ is the arm length of the detectors, $S_c(f)$ is the effective PSD of Galactic confusion noise, and $f_{\ast} = c/2\pi L_{\mathrm{arm}}$ is the transfer frequency that characterizes the frequency scale of the detector, corresponding to GWs whose period is comparable to the round-trip light-travel time along the detector arm  \citep{Cornish:2001bb}. Here $P_{\mathrm{IFO}}$ is modeled as \citep{2017arXiv170200786A, Robson:2018ifk}
\begin{equation}
\label{eq:LISAIFOPSDModel}
P_{\mathrm{IFO}}=\left(A_{\mathrm{IFO}}\right)^{2}\left(1+\left(\frac{2\rm{mHz}}{f}\right)^{4}\right) \mathrm{Hz}^{-1},
\end{equation}
where $A_{\mathrm{IFO}}$ is the IFO noise amplitude. The acceleration noise follows \citep{Larson:1999we, 2017arXiv170200786A, Robson:2018ifk}
\begin{equation}
\label{eq:LISAaccPSDModel}
P_{\mathrm{acc}}=\left(A_{\mathrm{acc}} \right)^{2}\left(1+\left(\frac{0.4\rm{mHz}}{f}\right)^{2}\right)\left(1+\left(\frac{f}{8\rm{mHz}}\right)^{4}\right) \mathrm{Hz}^{-1}
\end{equation}
where $A_{\mathrm{acc}}$ is the acceleration noise amplitude. The confusion noise is expressed as \citep{Cornish:2017vip, Robson:2018ifk}
\begin{equation}\label{eq:Sc}
S_c(f) = A f^{-7/3} e^{-f^\alpha + \beta f \sin{\kappa f}} \left[1+ \tanh(\gamma(f_k-f))\right] {\rm Hz^{-1}}.
\end{equation}
TianQin, with its shorter arm length, is sensitive to higher frequencies, allowing it to avoid galactic confusion noise \citep{Gong:2021gvw}. In contrast, LISA and Taiji are expected to face challenges from unresolved galactic binaries. For these two detectors, we adopt the confusion noise parameters $(A, \alpha, \beta, \kappa, \gamma, f_k ) = (9 \times 10^{-45}, 0.171, 292, 1020, 1680, 2.15 \rm ~mHz)$ for a one-year observation period \citep{Cornish:2017vip, Robson:2018ifk}. The detector response function $\mathcal{R}(f)$ ,which characterizes the frequency-dependent antenna pattern and polarization response of the interferometer, is computed using the analytical formulae derived in \citet{Zhang:2020khm}. For detector characterization, we adopt values (\(L_{\mathrm{arm}}\), \(A_{\mathrm{IFO}}\), \(A_{\mathrm{acc}}\)) of (\(2.5 \times 10^{6}~\mathrm{km}\), \(1.5 \times 10^{-11}~\mathrm{m}\), \(3.0 \times 10^{-15}~\mathrm{m~s^{-2}}\)) for LISA, (\(\sqrt{3} \times 10^{5}~\mathrm{km}\), \(1.0 \times 10^{-12}~\mathrm{m}\), \(1.0 \times 10^{-15}~\mathrm{m~s^{-2}}\)) for TianQin, and (\(3.0 \times 10^{6}~\mathrm{km}\), \(8.0 \times 10^{-12}~\mathrm{m}\), \(3.0 \times 10^{-15}~\mathrm{m~s^{-2}}\)) for Taiji, as specified in \citet{Gong:2021gvw}. In our S/N analysis of space-based interferometers, we utilize the numerical code \footnote{\href{https://github.com/yggong/transfer_function}{https://github.com/yggong/transfer$\_$function}} developed in the literature \citep{Robson:2018ifk, Liang:2019pry, Zhang:2020khm, Gong:2021gvw} to account for the distinct characteristics of LISA, Taiji, and TianQin.

\begin{figure}[htbp]
\includegraphics[width=1.0\linewidth]{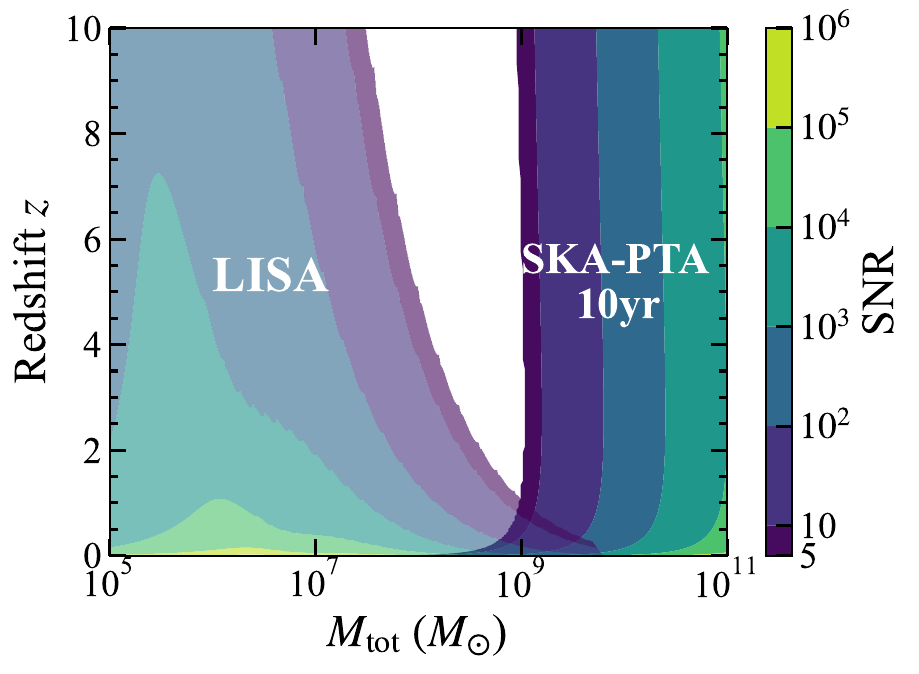} 
\caption{Contours of averaged S/N for LISA (left side) and SKA-PTA (right side) as functions of redshift and total binary mass $M_{\mathrm{tot}} = (1+q)M$, assuming equal-mass binaries ($q=1$). Shaded regions indicate detectable parameter space where the S/N exceeds the detection threshold ($\rho_{\mathrm{crit}} = 5$). For SKA-PTA, we consider an observational duration of $T_{\mathrm{obs}} = 10 $ yr with the fiducial parameter set ($\sigma_t, N_{\mathrm{pl}}, \Delta t$) = ($30$ ns, $500$, $0.02$ yr). \label{fig:S/N}}
\end{figure}

\begin{figure*}[htbp]
\begin{center}
  \begin{minipage}[b]{0.47\linewidth}
    \centering
    \includegraphics[keepaspectratio, scale=0.5]{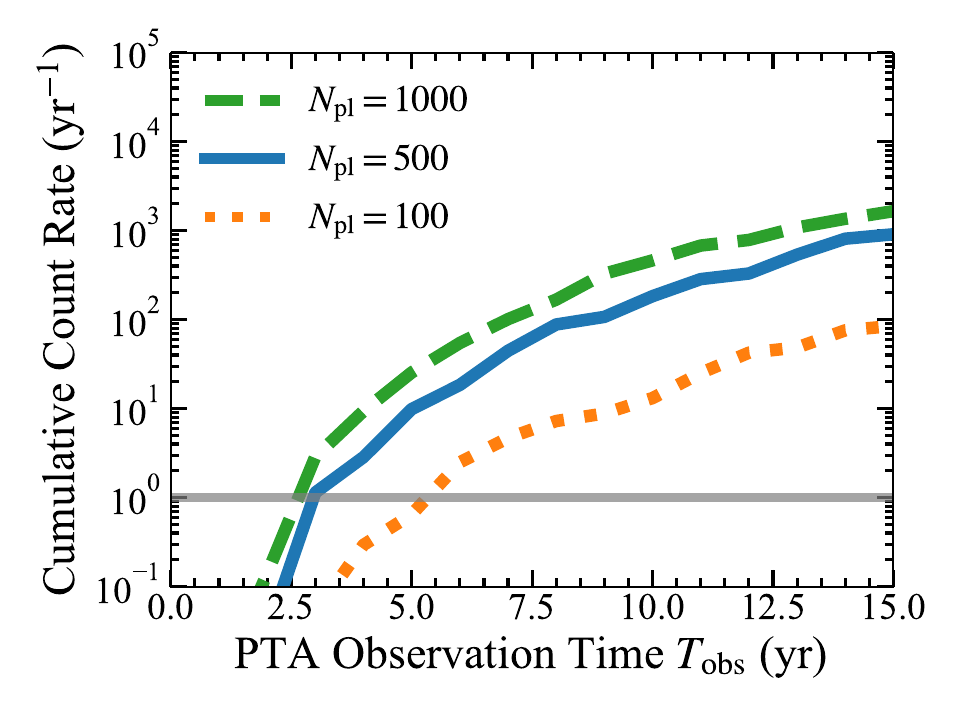}
  \end{minipage}
  \begin{minipage}[b]{0.47\linewidth}
    \centering
    \includegraphics[keepaspectratio, scale=0.5]{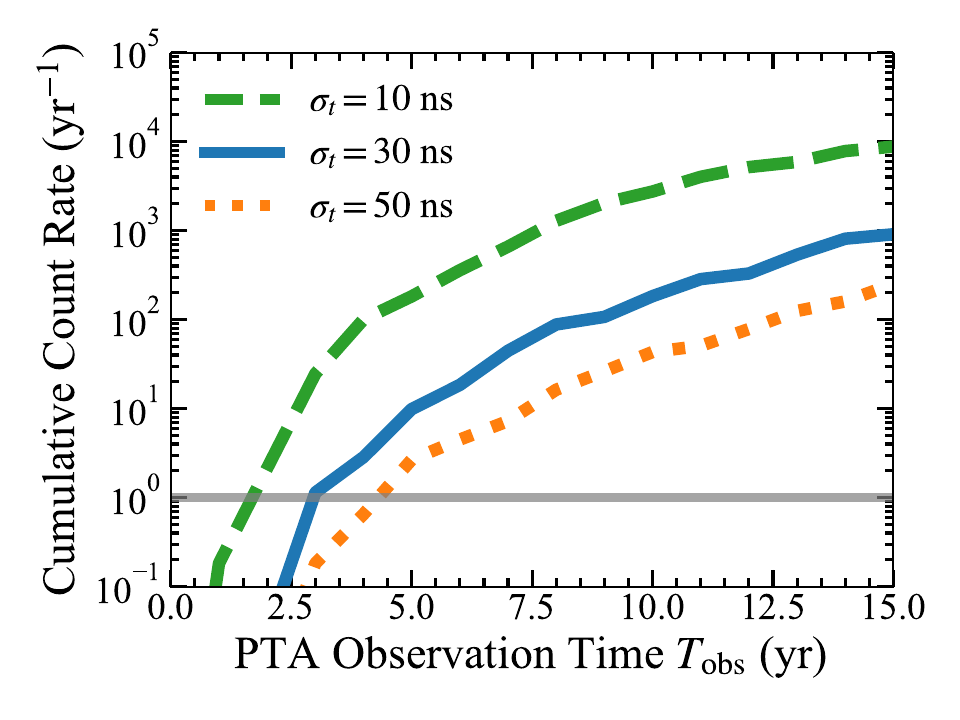}
  \end{minipage}
  \caption{The detection rate of detectable GW source counts as a function of PTA observation time. The solid-blue curve indicates the fiducial parameter case in both panels. The gray horizontal lines correspond to the source counts unity.
  \textit{Left}: Change the number of MSPs with $N_{\mathrm{pl}} = 1000$ (green dashed), 500 (blue solid), 100 (orange dotted), and fix $\sigma_t = 30$ ns. \textit{Right}: Change the RMS timing residuals with $\sigma_t = 10$ ns (green dashed), 30 ns (blue solid), 50 ns (orange dotted), and fix $N_{\mathrm{pl}} = 500$. \label{fig:PTA_cumulative}}
\end{center}
\end{figure*}

Figure~\ref{fig:S/N} illustrates signal-to-noise ratio contours for LISA and SKA-PTA as functions of redshift and the total binary mass $M_{\mathrm{tot}}$, assuming equal-mass binaries ($q = 1$). For space-borne detectors, we assume an observational duration of 4 years, while the SKA-PTA calculations assume a 10-year observational duration. Note that the S/N is calculated by integrating the waveform over the frequency evolution during this observation time \citep{Kaiser:2020tlg}. The contours demonstrate that LISA is primarily sensitive to SMBHBs with masses $\leq 10^8 M_{\odot}$ by expanding the observational window to redshifts $z \gtrsim 5$, while SKA-PTA will be sensitive to more massive systems in their extended inspiral phase. While current PTAs can only resolve GWs from very massive systems ($\sim 10^{10} M_{\odot}$), SKA-PTA will extend this detection capability down to $10^{8}$--$10^{9}\,M_{\odot}$.

\section{\label{sec:result} Analysis for Detectable GW Sources}
\subsection{\label{sec:SKA-PTA}Inspiraling GW Sources Detectable by SKA-PTA}
\begin{figure}[htbp]
\includegraphics[width=1.0\linewidth]{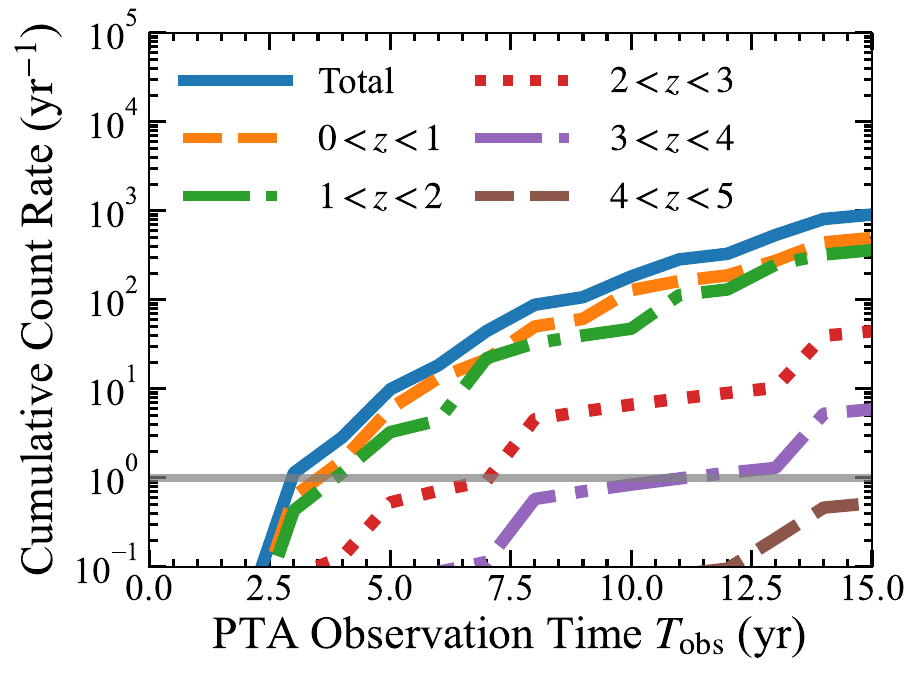} 
\caption{The detection number of GW sources per year from each redshift range, assuming fiducial parameter set. The dashed orange, dash-dotted green, dotted red, dash-dotted purple, dashed brown, and solid blue represent the number of sources at $0 < z < 1, 1 < z < 2, 2 < z < 3, 3 < z < 4, 4< z < 5$, and the total detection number, respectively. \label{fig:cum_count_z}}
\end{figure}
The detection rate of GW sources observed by SKA-PTA as a function of observational duration is shown in Figure~\ref{fig:PTA_cumulative}. We examine the sensitivity to key parameters by varying the number of monitored MSPs $N_{\mathrm{pl}}$ = (100, 500, 1000) and the RMS timing residual $\sigma_t$ = (10, 30, 50) ns around our fiducial configuration ($\sigma_t, N_{\mathrm{pl}}, \Delta t$) = (30 ns, 500, 0.02 yr), represented by the blue solid curve. SKA-PTA would achieve its first individual SMBHB detection within the initial few years of operation, as indicated by the intersection with the gray reference line. Over 10 years, the cumulative detection count is projected to reach $\sim 10^{2}$--$10^{3}$ systems, varying by up to an order of magnitude under different configurations. The left panel shows that increasing $N_{\mathrm{pl}}$ from 100 to 1000 enhances the 10-year count from $\sim 13$ to $\sim 460$ systems, representing approximately more than one order of magnitude improvement. Similarly, the right panel demonstrates that improving timing precision from $\sigma_t = 50$ ns to $\sigma_t = 10$ ns increases the detection rate by more than an order of magnitude, from $\sim 43$ to $\sim 2800$ systems over 10 years. These results underscore the critical importance of expanding the pulsar array size, achieving high timing precision, and extending the observational period to maximize SKA-PTA's discovery potential.

\begin{figure*}[htbp]
\begin{center}
  \begin{minipage}[b]{0.47\linewidth}
    \centering
    \includegraphics[keepaspectratio, scale=0.5]{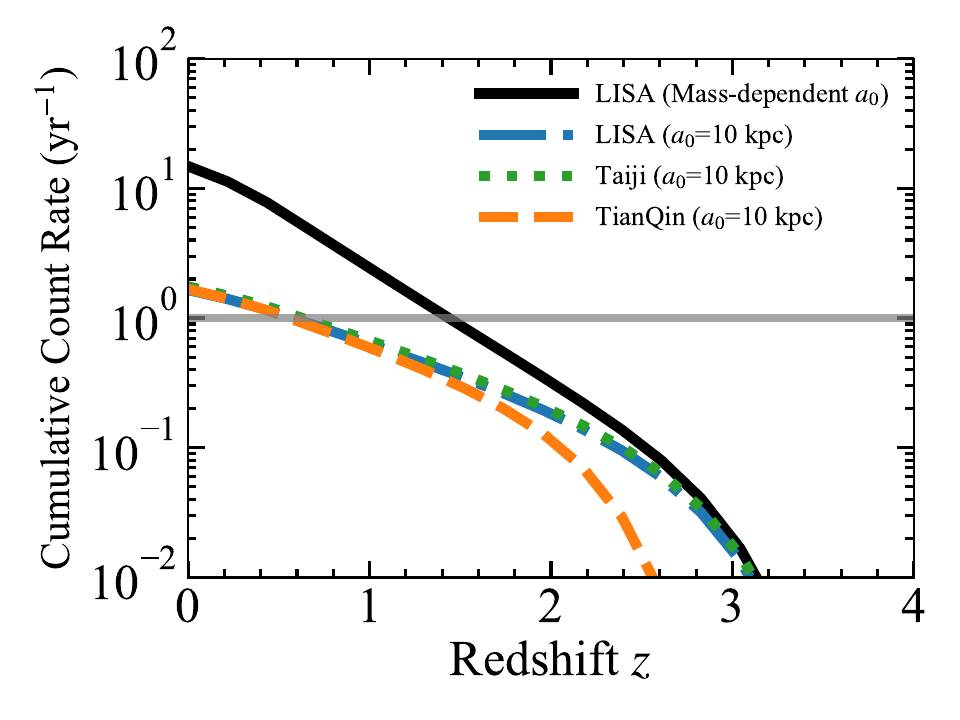}
  \end{minipage}
  \begin{minipage}[b]{0.47\linewidth}
    \centering
    \includegraphics[keepaspectratio, scale=0.5]{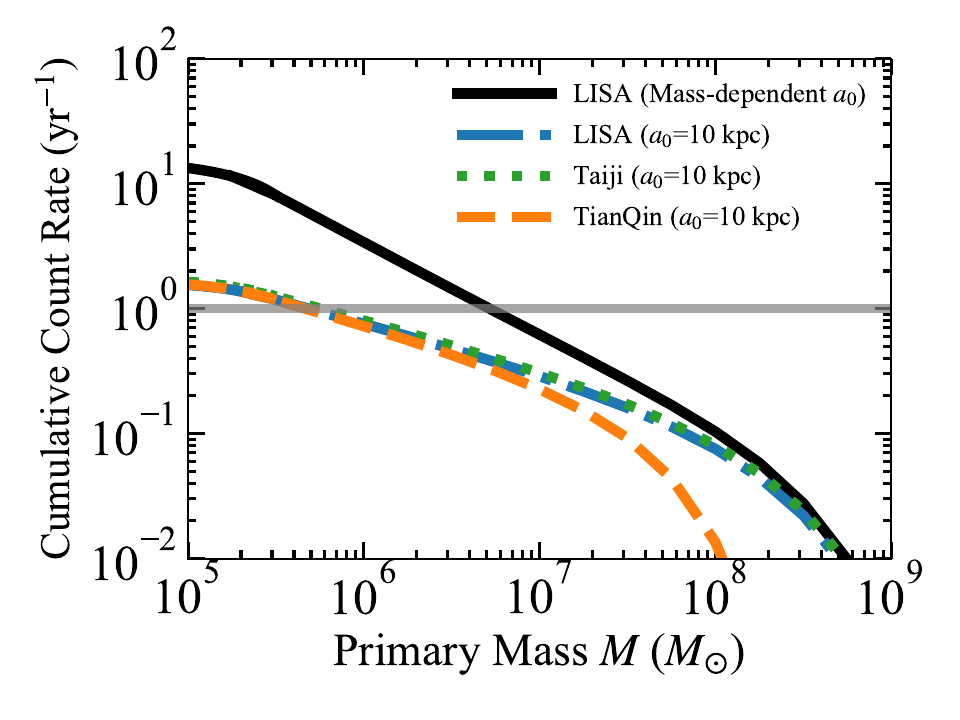}
  \end{minipage}
  \caption{\textit{Left}: Cumulative count rate per year as a function of the redshift for different space-borne GW observatories. The black solid curve indicates the LISA detection rate with mass-dependent separation prescription, while the dash-dotted blue, dashed-orange, and dotted-green curves represent the detection rate by LISA, TianQin, and Taiji assuming a fiducial initial separation case of $a_0$=10 kpc, respectively. The gray horizontal line corresponds to the count rate unity. \textit{Right}: Same as the left figure, but showing it as a function of the primary BH mass.
\label{fig:space_borne}}
\end{center}
\end{figure*}

Figure~\ref{fig:cum_count_z} presents the cumulative detection rate as a function of observational duration, decomposed into contributions from different redshift ranges. The majority of detectable sources ($\sim 70$ \%) reside at relatively low redshifts ($z < 1$) at the observational period of 10 yr, while the detection rate decreases progressively at higher redshifts. For instance, systems at $z > 2$ contribute only $\sim 4$ \% of the total detections at $T_{\rm obs} = 10$ yr. This redshift distribution reflects the declining number density of massive BHs at higher redshifts, as reflected in the black hole mass function (BHMF) derived from the AGN X-ray luminosity function \citep{Ueda:2014tma}.

\subsection{\label{subsec:space-borne}Annual Detection Rate for Space-Borne Interferometers}
The predicted annual detection rates for space-borne GW detectors are shown in Figure~\ref{fig:space_borne}. For SMBHBs with masses above $10^5~M_{\odot}$, our analysis predicts annual detection rates of approximately 1--2 systems per year across all three detector configurations, assuming a fiducial initial orbital separation of $a_0 = 10$ kpc. In contrast, the detection rates can be enhanced to $\sim$ 20 if we adopt the mass-dependent separation case. This sensitivity to the initial separation prescription reflects the dependence of SMBHB coalescence timescales on the starting point of binary evolution, particularly during the dynamical friction phase (see Section~\ref{sec:S/N}). Although these detectors are sensitive to high-redshift sources, most detections are expected in the local universe ($z \leq 1-2$), as illustrated in the left panel of Figure~\ref{fig:space_borne}. This concentration at low redshift results from the declining BHMF at $z \gtrsim 3$, as derived from the AGN X-ray luminosity function \citep{Ueda:2014tma}. Detection rates are dominated by lower-mass SMBHBs ($10^{5}$--$10^{6}~M_{\odot}$), with upper mass detection thresholds of approximately $10^9~M_{\odot}$ for LISA and Taiji, and $10^8~M_{\odot}$ for TianQin.

The three detectors exhibit distinct sensitivities due to their different arm lengths and orbital configurations. LISA and Taiji employ arm lengths of $L_{\mathrm{arm}} \sim 2.5 \times 10^6$ km and $L_{\mathrm{arm}} \sim 3.0 \times 10^6$ km, respectively \citep{Gong:2021gvw, LISA:2022yao, Chen:2023zkb}, while TianQin features a more compact configuration with $L_{\mathrm{arm}} \sim \sqrt{3} \times 10^5$ km \citep{TianQin:2015yph, Gong:2021gvw}. These configurations result in optimal sensitivity bands of $\sim 10^{-4}$--$10^{-1}$ Hz for LISA and Taiji, and ${\sim}10^{-2}$--$1$~Hz for TianQin \citep{Gong:2021gvw}. However, despite these instrumental differences, all three detectors are sensitive to remarkably similar SMBHB populations, with detectable sources consistently dominated by systems in the $10^{5}$--$10^{6}~M_{\odot}$ mass range at $z < 2$ across all configurations.

\section{\label{sec:discussion} Discussions} 
\subsection{Comparisons with Previous Works for SKA-PTA}\label{subsec:comparison_SKA}
Several previous works have investigated the detectability of individual SMBHBs with SKA-PTA using different methodological approaches \citep{Feng:2020nyw, Chen:2023hyj}. The primary distinction between our approach and these studies lies in the construction of the SMBHB population model. Previous works constructed their SMBHB populations based on local galaxy catalogs \citep{2017NatAs...1..886M} and galaxy stellar mass functions \citep{Bell:2003cj, Muzzin:2013yxa, 2019MNRAS.488.3143B} combined with galaxy merger rates inferred from the Illustris simulations \citep{Genel:2014lma, Rodriguez-Gomez:2015aua}, implicitly assuming that subsequent SMBH mergers occur following galaxy mergers. In contrast, our model directly incorporates observed dual AGN fractions combined with AGN X-ray luminosity functions, providing independent observational constraints on the SMBHB population.

We first compare our predictions with \citet{Feng:2020nyw}, who demonstrated that SKA-PTA could achieve substantial SMBHB detections even with a relatively modest array configuration. They predicted that a configuration with $N_{\mathrm{pl}} = 20$ pulsars combined with high timing precision ($\sigma_t = 20$ ns) and frequent observations ($1/\Delta t = 20~\mathrm{yr}^{-1}$) could achieve first detection within approximately 5 years and reach detection rates exceeding 100 SMBHBs after about 10 years. Adopting the same PTA configuration, our model predicts approximately $8\times10^{-1}$ sources after 10 years, roughly two orders of magnitude lower than their prediction. This difference primarily reflects the underlying SMBHB formation rate: our model employs the dual AGN fraction from \citet{Kusakabe:2025rmn}, which shows a declining trend toward higher black hole masses (dual AGN fraction of $\leq 0.1\%$ for $\geq 10^9~M_{\odot}$ binaries), resulting in suppressed formation rates for the most massive binaries that dominate the PTA sensitivity range.

Similarly, \citet{Chen:2023hyj} conducted a comprehensive study on PTA detectability, exploring two parameter configurations for SKA-PTA: a conservative case with ($\sigma_t, N_{\mathrm{pl}}, \Delta t$) = (100 ns, 100, 0.04 yr) and an optimistic case with ($\sigma_t, N_{\mathrm{pl}}, \Delta t$) = (20 ns, 1000, 0.02 yr), predicting approximately $10^2$ and $5 \times 10^4$ detectable sources after 10 years, respectively. Adopting these same parameter configurations, our model predicts approximately $6 \times 10^{-2}$ and $1 \times 10^3$ sources respectively, showing underestimation by two to three orders of magnitude. This systematic difference across different array configurations indicates a fundamental distinction in the underlying SMBHB population models, particularly regarding the SMBHB formation rate discussed above. Thus, future SKA-PTA observations of individual sources will be pivotal for deciphering the high-mass end of the SMBHB population synthesis.

\subsection{Comparisons with Previous Works for Space-Borne Detectors}\label{subsec:comparison_SB}
In the literature, predictions for the detectability of space-borne interferometers have been made employing SAMs or cosmological hydrodynamical simulations \citep{Sesana:2010wy, Klein:2015hvg, Salcido:2016oor, Bonetti:2018tpf, 2019MNRAS.486.2336D, Katz:2019qlu, Barausse:2020mdt, Curylo:2021pvf, 2024A&A...686A.183I}. In the semi-analytical approach, the evolution of galaxies and their central BHs is modeled using analytic prescriptions for processes such as galaxy mergers, black hole seeding, accretion, and binary hardening, implemented within merger trees that trace the hierarchical assembly of dark matter halos \citep{Sesana:2010wy, Klein:2015hvg, Bonetti:2018tpf, 2019MNRAS.486.2336D, Barausse:2020mdt, Curylo:2021pvf, 2024A&A...686A.183I}. Hydrodynamical simulations, in contrast, trace SMBH coalescence in cosmological volumes, with subsequent binary evolution toward merger typically supplemented by a constant delay based on the gas abundance of host galaxies \citep[e.g.,][]{Salcido:2016oor} or subgrid prescriptions for unresolved dynamical processes such as dynamical friction, stellar hardening, the viscous torque from a gas disk, and the GW driven inspiral \citep[e.g.,][]{Katz:2019qlu}. Overall, these studies generally predict LISA detection rates spanning from a few to approximately 100 events per year, with model-dependent uncertainties arising from the SMBHB population model or the treatment of SMBHB orbital evolution.

Our predictions of 1–20 events per year, depending on orbital evolution prescriptions, fall within the ranges forecasted by previous works. Thus, our model confirms previous theoretical predictions through an independent observational approach. However, uncertainties remain, primarily originating from the BHMF at intermediate masses ($10^5$--$10^7\,M_{\odot}$). Due to the absence of direct observational constraints in this regime, we construct our population model by extrapolating the BHMF at masses exceeding $10^7\,M_{\odot}$ \citep{Ueda:2014tma}. However, recent JWST observations have significantly extended the BHMF down to the lower-mass regime, identifying BHs with masses as low as $\sim 10^{5-6} M_{\odot}$ residing at high-$z$ Universe of $z \geq 4-5$ \citep{2024ApJ...968...38K, 2025ApJ...986..165T, Fei:2025hzd}. Additionally, these observations have revealed that the number density of such faint AGNs is substantially higher than previously estimated, exceeding pre-JWST extrapolations by a factor of $\sim 10$ \citep{Matthee:2023utn, 2025ApJ...986..165T}. By calibrating AGN population with these constraints, recent studies \citep{Liu:2024kig, 2025arXiv251120414C} predict coalescence rates of a few dozen per year at $z \ge 5$ with BH masses exceeding $10^{3-4} M_{\odot}$. These findings indicate that if models are calibrated against these recent JWST results, there is a possibility of detecting a significant number of GW sources even at such high redshifts. Encouragingly, future observational advances will offer promising avenues for improvement. Dynamical measurements in nearby galaxies, which measure BH masses by modeling the motions of surrounding stars or gas under its gravitational influence, provide the most direct constraints on BH masses in this range \citep{2012AdAst2012E..15N, 2019ApJ...872..104N}. Future observations with extremely large telescopes will extend these measurements to larger volumes and lower masses \citep{2020ARA&A..58..257G}. Ultraluminous X-ray sources, especially those with X-ray luminosities $L_{X} \geq 10^{41}$~erg~s$^{-1}$ have emerged as potential tracers of intermediate-mass BHs, with several strong candidates identified through their distinctive spectral and timing properties \citep{Kaaret:2017tcn, Barrows:2019pxt}. Additionally, tidal disruption events provide independent constraints on the intermediate-mass BH population, particularly in dwarf galaxies and globular clusters where lower-mass BHs are expected to reside \citep{2020ARA&A..58..257G, Melchor:2025fey}. Combining these diverse observational probes will enable the construction of more robust mass functions extending to intermediate mass regimes, improving detection rate predictions for space-borne interferometers.

\subsection{Impact of Inhibiting Noise Components for Pulsar Timing Detection}\label{subsec:red_noise}
The successful detection of GWs in the nHz regime critically depends on the ability to effectively process and mitigate various noise components in PTA observations. Regarding noise treatment in nHz GW measurements, we assume white Gaussian noise and neglect red noise contributions as described in Section~\ref{sec:threshold}. Understanding and properly accounting for these noise contributions is particularly crucial since they may mask continuous GW signals from individual binaries. This is especially important in the low-frequency domain, where many promising SMBHB sources are expected to reside \citep{Feng:2020nyw}. Including these noise sources, the full PSD can be expressed as \citep{Thrane:2013oya, Lam:2018uta}:
\begin{equation}
P_{n}(f) = P_{\rm SGWB}(f) + \sum_{i}^{N_{\rm pl}}P_{i}(f) 
\end{equation}
where the PSD of the SGWB is modeled as a single power-law \citep{Jenet:2006sv}
\begin{equation}
P_{\rm SGWB}(f) = \frac{A_{\rm SGWB}^{2}}{12 \pi^{2}}\left(\frac{f}{f_{\rm year}}\right)^{2
\alpha^{\prime}}f^{-3} 
\end{equation}
with $A_{\rm SGWB}$ as the strain amplitude at a frequency of $f_{\mathrm{year}} = 1\ \rm yr^{-1}$ and $\alpha^{\prime}= -2/3$ is the spectral index of the characteristic strain. The PSD for each pulsar can be rewritten from Eq.~\ref{eq:WNPSD} as
\begin{equation}
\label{eq:PulsarPSD}
P_{i}(f)  =  P_{{\rm RN},i}(f) + P_{{\rm WN},i} 
\end{equation}
where the red noise contribution is modeled by a single power-law \citep{Hazboun:2019vhv}
\begin{equation}
P_{{\rm RN},i}(f) = A_{{\rm RN},i} \left(\frac{f}{f_{\rm year}}\right)^{-\gamma_{\mathrm{RN},i}} , \hspace{5mm} \gamma_{i} > 0
\end{equation}
where $A_{{\rm RN},i}$ is the pulsar red noise amplitude and $\gamma_{\mathrm{RN},i}$ is the spectral index of the red noise power.

\begin{figure}[htbp]
\includegraphics[width=1.0\linewidth]{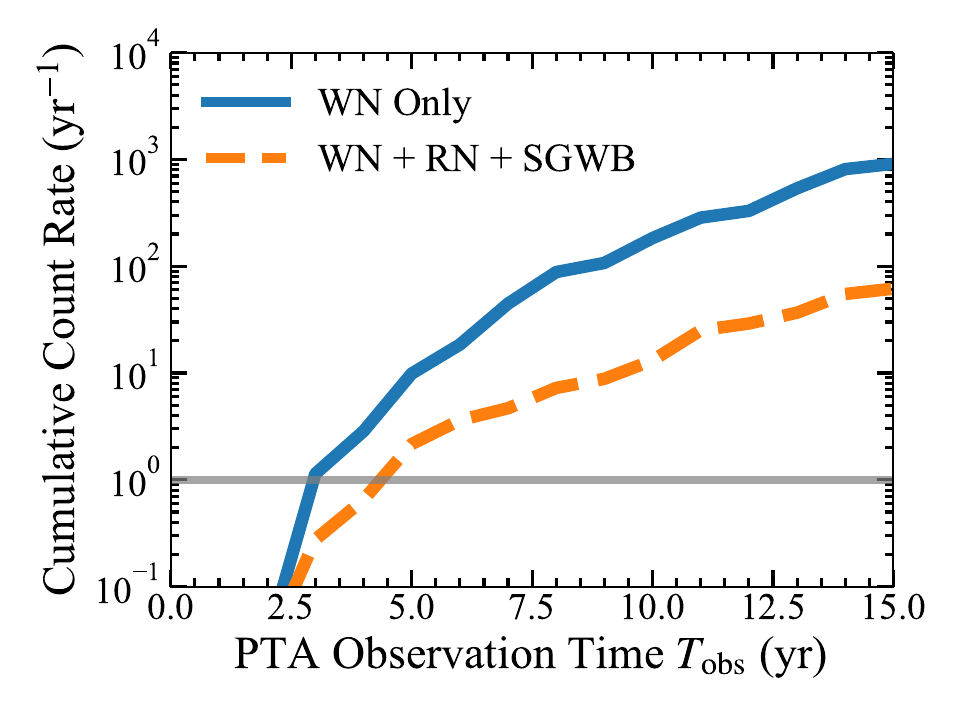} 
\caption{The detection number of GW sources per year against different noise configurations. The blue solid curve indicates the noise model, which only contains white noise, the same as in Figure~\ref{fig:PTA_cumulative} (WN Only; our baseline model), while the orange dashed curve denotes the noise containing sampled red noise and SGWB in addition to white noise (WN + RN + SGWB). Here $A_{\mathrm{SGWB}} = 2.4 \times 10^{-15}$ \citep{NANOGrav:2023gor} and $\mathrm{log_{10}}A_{{\rm RN},i} \in [-17,-13]$ ,$\gamma_{\mathrm{RN},i} \in [1,5]$ \citep{Babak:2024yhu} are adopted for SGWB and red noise, respectively.
\label{fig:WN_RN_GWB}}
\end{figure}

Figure~\ref{fig:WN_RN_GWB} compares the white-noise-only model (WN-Only: our fiducial case) with one including sampled red noise and SGWB (WN + RN + SGWB). We adopt an SGWB amplitude $A_{\mathrm{SGWB}} = 2.4 \times 10^{-15}$, corresponding to the GW strain amplitude estimated from NANOGrav 15 yr data set \citep{NANOGrav:2023gor}. For the red noise parameters of each pulsar, we assume that the parameters are uniformly distributed as $\mathrm{log_{10}}A_{{\rm RN},i} \in [-17,-13]$ and $\gamma_{\mathrm{RN},i} \in [1,5]$ \citep{Babak:2024yhu}. As a result, these noise components disturb the highly accurate observational data accumulation and decrease the detectable number of GW sources by approximately one order of magnitude in the range considered here. Additional unmodeled noise sources include temporal variations in dispersion measure and scattering \citep{Babak:2024yhu}. These timing residual noise components substantially reduce low-frequency GW detection probability \citep{Feng:2020nyw}.

Identifying individual SMBHB signals in the presence of these noise components requires sophisticated signal-extraction techniques. Fully Bayesian pipelines simultaneously model timing parameters, multiple noise components (white noise, red noise, dispersion variations), SGWB, and potential continuous wave signals, as implemented in frameworks such as the PTA data analysis suite \href{http://doi.org/10.5281/zenodo.4059815}{\texttt{enterprise}}. Recent targeted searches for individual SMBHBs in current PTA datasets illustrate both the challenges and potential: targeted searches in EPTA \citep{EPTA:2023gyr} and PPTA \citep{Zhao:2025pgg} data have yielded upper limits, confirming that while current PTAs have found evidence of SGWB detection, individual SMBHB detections remain just beyond reach with current sensitivity. However, \citet{Agarwal:2025cag} applied Bayesian model comparison to NANOGrav 15-year data and identified two marginal SMBHB candidates, demonstrating that we are approaching the threshold of individual source detection and illustrating how refined noise models can enhance detection confidence. Similarly, recent simulations utilizing the $\mathcal{F}$-statistic—a frequentist maximum likelihood method—have demonstrated that by incorporating the SGWB into the noise covariance matrix, individual SMBHBs can be effectively resolved even in regimes where the SGWB exceeds the white noise level \citep{Furusawa:2025yuc}. The transition to resolved sources with SKA-PTA will depend on enhanced timing precision through improved instrumentation, expansion of pulsar arrays for increased sensitivity, and refined noise characterization through extended observational baselines. These combined improvements will position SKA-PTA to achieve the conclusive detections of individual SMBHBs within the coming decade.

\subsection{\label{subsec:S/N}Signal-to-Noise Ratio Distribution for GW Sources}
\begin{figure}[htbp]
\includegraphics[width=1.0\linewidth]{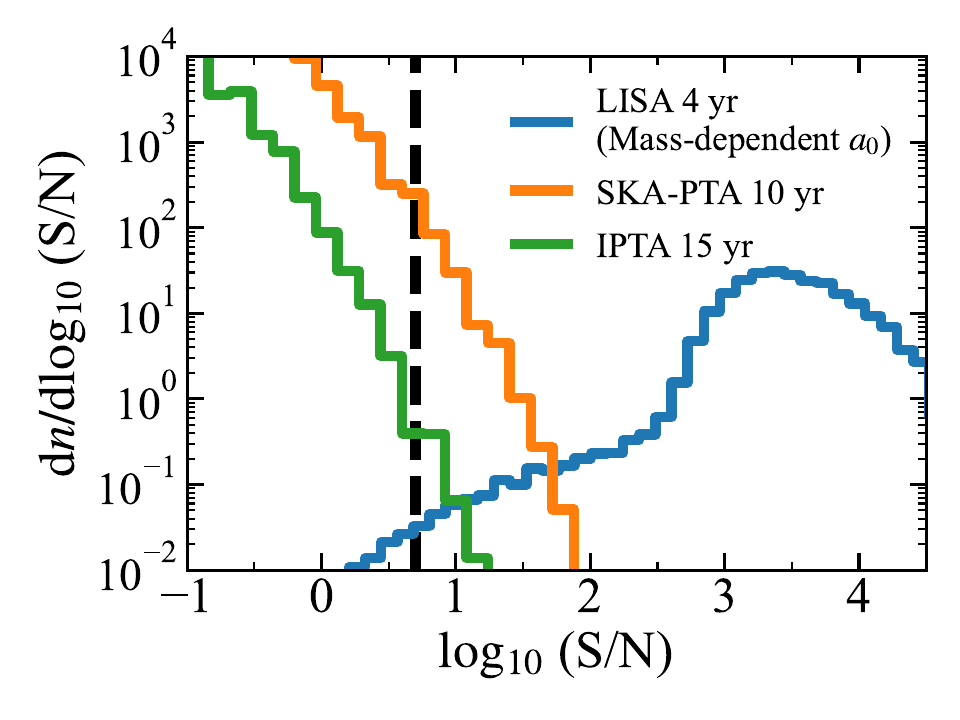} 
\caption{S/N distribution for LISA 4 yr (blue), SKA-PTA 10 yr (orange), and IPTA 15 yr (green) observation, respectively. For SKA-PTA, the parameters are assumed to be fiducial as ($\sigma_t, N_{\mathrm{pl}}, \Delta t$) = ($30$ ns, $500$, $0.02$ yr). In case of IPTA, ($\sigma_t, N_{\mathrm{pl}}, \Delta t$) = ($100$ ns, $100$, $0.04$ yr) are adopted with observational period $T_{\mathrm{obs}} =15$ yr. All the PTA noise models are based on the WN-Only described in Section \ref{subsec:red_noise}. S/N distribution of LISA is shown as the cumulative counts of 4-year observations assuming the mass-dependent $a_0$ model. The vertical dashed line marks the nominal detection threshold of S/N =5. \label{fig:SNR_distribution}}
\end{figure}

We now present a comprehensive analysis of the anticipated S/N distributions for detectable SMBHB populations across each GW detector in Figure~\ref{fig:SNR_distribution}. For SKA-PTA, we show predictions for 10-year with our fiducial parameters, alongside the current International PTA (IPTA) capability of ($\sigma_t, N_{\mathrm{pl}}, \Delta t$) = ($100$ ns, $100$, $0.04$ yr) as a reference during 15 yr observations \citep{Antoniadis:2022pcn, NANOGrav:2023hde, EPTA:2023sfo, Reardon:2023zen}. For space-borne detectors, we present results for a 4-year LISA mission assuming the mass-dependent separation model. 

As for the PTA detections, Figure~\ref{fig:SNR_distribution} reveals distinct characteristics across different observational PTA capabilities. With current IPTA configurations, the vast majority of potentially detectable sources fall near the detection threshold of S/N $\sim$ 5, with only a handful of sources ($\sim$ 1) that could be captured. This marginal detectability explains why individual SMBHB detections remain elusive despite the evidence of SGWB existence \citep{NANOGrav:2023gor, Reardon:2023gzh, EPTA:2023xxk}. In contrast, SKA-PTA dramatically improves the detection landscape. After 10 years of observation, the cumulative number of detectable sources increases substantially with several hundred sources above the detection threshold, though many remain in the relatively low S/N regime ($\log_{10}(\mathrm{S/N}) \sim 1$). This underscores the critical importance of sophisticated data analysis techniques for effectively extracting continuous GW signals, as discussed in Section~\ref{subsec:red_noise}. The ability to accurately model and subtract noise components will be paramount for maximizing the scientific yield from these numerous detections.

Space-borne detectors present a complementary picture. LISA's 4-year mission is expected to detect fewer sources compared to SKA-PTA accumulations, but with substantially higher significance levels. As shown in Figure~\ref{fig:SNR_distribution}, the LISA-detectable population exhibits S/N values typically reaching $\log_{10}(\mathrm{S/N}) \geq 2$-$3$, or even higher. A notable feature is the pronounced peak around $\log_{10}(\mathrm{S/N}) \sim 3$, which corresponds to the optimal sensitivity regime of LISA in the mHz frequency band, where the detector exhibits maximal sensitivity for SMBHBs in the $10^5$--$10^6$ $M_\odot$ mass range (See~Fig.~\ref{fig:S/N}). For sources within this mass range, these high signal-to-noise detections will enable precise parameter estimation, including BH masses, spins, orbital parameters, and sky localization \citep{Baibhav:2020tma, Ruan:2019tje}. Although our model does not extend below $10^5\,M_\odot$, we note that the existence of a substantial lower-mass population (e.g., $10^3$--$10^4\,M_\odot$) would preferentially populate the low-S/N tail, with some potentially residing near LISA's detection threshold. Depending on their number density, such systems could increase the source counts in the lower-S/N bins, making our source count predictions conservative.

This distinct difference in S/N distributions can be understood by comparing the detector sensitivities with the source population properties shown in Figure~\ref{fig:S/N}. Fundamentally, LISA observes SMBHBs during their final inspiral and merger phases where GW emission is strongest, while PTAs detect systems in earlier, extended inspiral phases with weaker signals, leading to systematically higher S/N for LISA detections. Moreover, for space-borne detectors, the primary targets ($10^{5-7} M_{\odot}$) remain detectable with high significance even at high redshifts, leaving few events near the detection threshold. In contrast, while PTAs can detect super-massive systems ($> 10^9 M_{\odot}$) at cosmological distances, such sources are intrinsically rare. The PTA population is instead dominated by systems with $\lesssim 10^9 M_{\odot}$, which are resolvable only at lower redshifts ($z \lesssim 1$). Consequently, a large population of these sources accumulates near the detection limit, resulting in a distribution skewed toward lower S/N.

The stark contrast in S/N distributions between PTA and space-borne detectors—with PTAs observing many sources at moderate significance and space-borne detectors observing sources at high significance—highlights their complementary roles in the emerging multi-band GW astronomy landscape. Together, these facilities will probe SMBHB populations across different mass ranges, redshifts, and evolutionary stages, providing a comprehensive view of SMBH binary evolution.

\subsection{\label{subsec:Prospects} Prospects for GW Probes and Implications for Multi-messenger Astronomy}
The detection rate predictions presented in this work establish concrete expectations for the observational capabilities of next-generation GW facilities. These predictions, grounded in observed dual AGN fractions and AGN X-ray luminosity functions, provide a framework for interpreting forthcoming observations and planning targeted follow-up campaigns. Building on these quantitative forecasts, we now discuss the implications for multi-band GW astronomy and multi-messenger astronomy.

Early SKA-PTA observations will be pivotal in distinguishing the detected nHz SGWB \citep{NANOGrav:2023gor, Reardon:2023gzh, EPTA:2023xxk} from alternative cosmological sources, such as first-order phase transitions \citep{Caprini:2015zlo, Caprini:2019egz, Hindmarsh:2020hop}, scalar-induced GWs \citep{Domenech:2021ztg, Yuan:2021qgz}, and cosmic strings or domain walls \citep{Vilenkin:1984ib, Hindmarsh:1994re, Saikawa:2017hiv}. Our predictions suggest that SKA-PTA will detect its first individual SMBHB within a few years of operation, providing direct confirmation that SMBHBs contribute significantly to the observed background. In addition, the accumulation of hundreds to thousands of individual detections over a decade will enable statistical tests of SMBHB population models and constrain the contribution of astrophysical binaries to the total SGWB. 

Complementing PTA observations, space-borne interferometers will enable detailed studies of SMBHBs during their final inspiral and merger phases. These measurements will constrain orbital dynamics influenced by stellar and gaseous environments \citep{Burke-Spolaor:2018bvk}, clarifying orbital decay mechanisms and merger histories within galactic nuclei. The multi-detector network will significantly enhance sky localization (potentially improving precision by up to four orders of magnitude, reaching $\lesssim$ 0.01 deg$^2$ for the brightest sources at low redshifts) and parameter estimation accuracy—including luminosity distance, masses, and spins—through complementary sensitivity profiles and extended baselines \citep{Ruan:2019tje, Ruan:2020smc, Cai:2023ywp}.

The large number of SMBHB detections predicted by our model, particularly from SKA-PTA, will create unprecedented opportunities for multi-messenger follow-up campaigns. The Vera Rubin Observatory's Legacy Survey of Space and Time (LSST) \citep{LSST:2008ijt} is expected to identify AGNs exhibiting periodic variability or other binary signatures \citep{Kelley:2018fur, Xin:2021mmk, Xin:2024fci, Chiesa:2025gwk, Xin:2025voy}. Cross-correlating these electromagnetic candidates with PTA data will enable systematic identification of SMBHB systems, dramatically improving GW source localization. In addition, next-generation extremely large telescopes—including the Thirty Meter Telescope \citep{TMTInternationalScienceDevelopmentTeamsTMTScienceAdvisoryCommittee:2015pvw}, Giant Magellan Telescope \citep{2012SPIE.8444E..1HJ}, and European Extremely Large Telescope \citep{Padovani:2023dxc}—will provide high-resolution spectroscopy to characterize host galaxy properties, stellar populations, gas dynamics, and binary environments. In parallel, high-resolution radio interferometry will play a dual role in characterizing these systems. Precise pulsar distance measurements via the SKA or VLBI can significantly improve the GW source localizatio \citep{Kato:2023tfz, Kato:2025set}, drastically reducing the search area and enabling the identification of host galaxies. Furthermore, instruments like the Next-Generation Very Large Array (ngVLA) or sub-mm VLBI will resolve dual active nuclei at pc-scale separations \citep{Burke-Spolaor:2018ghm, DOrazio:2017dyb, Zhao:2023kff}, providing direct confirmation of binary candidates and constraints on circumbinary disk structures.

Together, this multi-band, multi-messenger approach will enable end-to-end tracking of SMBHBs from kpc-scale separations through the nHz regime monitored by PTAs to final coalescence in the mHz band by space-borne interferometers. This comprehensive observational coverage will address fundamental questions about SMBHB formation efficiency, orbital evolution timescales, environmental influences on merger rates, and the role of gas and stellar dynamics. The detection rate predictions presented in this work provide a quantitative foundation for planning these multi-messenger campaigns and interpreting the forthcoming wealth of GW observations.

\section{\label{sec:conclusion} Conclusions}    
This work provides a comprehensive framework for predicting the detection prospects of SMBHBs captured by future GW observatories. Unlike previous studies that primarily rely on semi-analytical models or cosmological simulations, our approach directly constrains SMBHB populations using observed dual AGN fractions combined with AGN X-ray luminosity functions. Through systematic signal-to-noise ratio analysis, we evaluate the detection capabilities of both space-borne interferometers (LISA, Taiji, TianQin) and next-generation PTA (SKA-PTA).

Our analysis reveals distinct but complementary detection capabilities across different GW observatories. For SKA-PTA, we predict the first individual SMBHB detection within a few years of operation, with detection rates of $\sim$100-1000 systems after approximately 10 years of observation, depending on array configuration. This facility will be primarily sensitive to the most massive binary systems ($10^8-10^9\,M_{\odot}$) during their extended inspiral phases, with most detections concentrated at relatively low redshifts ($z \lesssim 1$). However, our analysis incorporating inhibiting components such as pulsar red noise and SGWB contributions indicates that detection counts can be reduced by one order of magnitude compared to idealized white noise scenarios. This emphasizes the critical importance of sophisticated data analysis techniques for extracting continuous GW signals from complex noise components.

Space-borne detectors present a complementary picture, predicted to capture approximately 1--20 SMBHB coalescence events annually, depending on assumptions about the orbital evolution prescriptions. These detections will be concentrated in the relatively local universe ($z \lesssim 2$) and dominated by lower-mass systems ($\lesssim 10^{6}\,M_{\odot}$), with typical signal-to-noise ratios substantially higher ($\log_{10}(\mathrm{S/N}) \gtrsim 2$--$3$), enabling precise binary parameter estimation and sky localization.

These facilities will open unprecedented observational windows into SMBHB characterization across complementary frequency bands. The detection rate predictions presented here provide quantitative benchmarks for interpreting upcoming GW observations. \\

\section*{Acknowledgements}
YI is supported by NAOJ ALMA Scientific Research Grant Number 2021-17A; JSPS KAKENHI Grant Number JP18H05458, JP19K14772, and JP22K18277; and World Premier International Research Center Initiative (WPI), MEXT, Japan. DT was supported in part by the JSPS Grant-in-Aid for Scientific Research (22K21349). KT is partially supported by JSPS KAKENHI Grant Numbers 20H00180, 21H01130, 21H04467, and 24H01813, and Bilateral Joint Research Projects of JSPS.

\bibliography{references}

@ARTICLE{1990ApJ...361..300F,
       author = {{Foster}, R.~S. and {Backer}, D.~C.},
        title = "{Constructing a Pulsar Timing Array}",
      journal = {\apj},
         year = 1990,
        month = sep,
       volume = {361},
        pages = {300},
          doi = {10.1086/169195},
       adsurl = {https://ui.adsabs.harvard.edu/abs/1990ApJ...361..300F}
}

@article{DOrazio:2023rvl,
    author = "D'Orazio, Daniel J. and Charisi, Maria",
    title = "{Observational Signatures of Supermassive Black Hole Binaries}",
    journal = "arXiv e-prints",
    eprint = "2310.16896",
    archivePrefix = "arXiv",
    primaryClass = "astro-ph.HE",
    year = "2023",
    month = "10",
    note = "arXiv:2310.16896"
}

@article{LISA:2022yao,
    author = "Seoane, Pau Amaro and others",
    collaboration = "LISA",
    title = "{Astrophysics with the Laser Interferometer Space Antenna}",
    eprint = "2203.06016",
    archivePrefix = "arXiv",
    primaryClass = "gr-qc",
    doi = "10.1007/s41114-022-00041-y",
    journal = "Living Rev. Rel.",
    volume = "26",
    number = "1",
    pages = "2",
    year = "2023"
}

@article{Feng:2020nyw,
    author = "Feng, Yi and Li, Di and Zheng, Zheng and Tsai, Chao-Wei",
    title = "{Supermassive Binary Black Hole Evolution can be traced by a small SKA Pulsar Timing Array}",
    eprint = "2005.11118",
    archivePrefix = "arXiv",
    primaryClass = "astro-ph.IM",
    doi = "10.1103/PhysRevD.102.023014",
    journal = "Phys. Rev. D",
    volume = "102",
    number = "2",
    pages = "023014",
    year = "2020"
}

@article{Burke-Spolaor:2018bvk,
    author = "Burke-Spolaor, Sarah and others",
    title = "{The Astrophysics of Nanohertz Gravitational Waves}",
    eprint = "1811.08826",
    archivePrefix = "arXiv",
    primaryClass = "astro-ph.HE",
    doi = "10.1007/s00159-019-0115-7",
    journal = "Astron. Astrophys. Rev.",
    volume = "27",
    number = "1",
    pages = "5",
    year = "2019"
}

@article{NANOGrav:2023hde,
    author = "Agazie, Gabriella and others",
    collaboration = "NANOGrav",
    title = "{The NANOGrav 15 yr Data Set: Observations and Timing of 68 Millisecond Pulsars}",
    eprint = "2306.16217",
    archivePrefix = "arXiv",
    primaryClass = "astro-ph.HE",
    doi = "10.3847/2041-8213/acda9a",
    journal = "Astrophys. J. Lett.",
    volume = "951",
    number = "1",
    pages = "L9",
    year = "2023"
}

@article{NANOGrav:2023gor,
    author = "Agazie, Gabriella and others",
    collaboration = "NANOGrav",
    title = "{The NANOGrav 15 yr Data Set: Evidence for a Gravitational-wave Background}",
    eprint = "2306.16213",
    archivePrefix = "arXiv",
    primaryClass = "astro-ph.HE",
    doi = "10.3847/2041-8213/acdac6",
    journal = "Astrophys. J. Lett.",
    volume = "951",
    number = "1",
    pages = "L8",
    year = "2023"
}

@article{EPTA:2023sfo,
    author = "Antoniadis, J. and others",
    collaboration = "EPTA",
    title = "{The second data release from the European Pulsar Timing Array - I. The dataset and timing analysis}",
    eprint = "2306.16224",
    archivePrefix = "arXiv",
    primaryClass = "astro-ph.HE",
    doi = "10.1051/0004-6361/202346841",
    journal = "Astron. Astrophys.",
    volume = "678",
    pages = "A48",
    year = "2023"
}

@article{EPTA:2023xxk,
    author = "Antoniadis, J. and others",
    collaboration = "EPTA, InPTA",
    title = "{The second data release from the European Pulsar Timing Array - IV. Implications for massive black holes, dark matter, and the early Universe}",
    eprint = "2306.16227",
    archivePrefix = "arXiv",
    primaryClass = "astro-ph.CO",
    doi = "10.1051/0004-6361/202347433",
    journal = "Astron. Astrophys.",
    volume = "685",
    pages = "A94",
    year = "2024"
}

@article{Zic:2023gta,
    author = "Zic, Andrew and others",
    title = "{The Parkes Pulsar Timing Array third data release}",
    eprint = "2306.16230",
    archivePrefix = "arXiv",
    primaryClass = "astro-ph.HE",
    doi = "10.1017/pasa.2023.36",
    journal = "Publ. Astron. Soc. Austral.",
    volume = "40",
    pages = "e049",
    year = "2023"
}

@article{Reardon:2023zen,
    author = "Reardon, Daniel J. and others",
    title = "{The Gravitational-wave Background Null Hypothesis: Characterizing Noise in Millisecond Pulsar Arrival Times with the Parkes Pulsar Timing Array}",
    eprint = "2306.16229",
    archivePrefix = "arXiv",
    primaryClass = "astro-ph.HE",
    doi = "10.3847/2041-8213/acdd03",
    journal = "Astrophys. J. Lett.",
    volume = "951",
    number = "1",
    pages = "L7",
    year = "2023"
}

@article{Reardon:2023gzh,
    author = "Reardon, Daniel J. and others",
    title = "{Search for an Isotropic Gravitational-wave Background with the Parkes Pulsar Timing Array}",
    eprint = "2306.16215",
    archivePrefix = "arXiv",
    primaryClass = "astro-ph.HE",
    doi = "10.3847/2041-8213/acdd02",
    journal = "Astrophys. J. Lett.",
    volume = "951",
    number = "1",
    pages = "L6",
    year = "2023"
}

@article{Xu:2023wog,
    author = "Xu, Heng and others",
    title = "{Searching for the Nano-Hertz Stochastic Gravitational Wave Background with the Chinese Pulsar Timing Array Data Release I}",
    eprint = "2306.16216",
    archivePrefix = "arXiv",
    primaryClass = "astro-ph.HE",
    doi = "10.1088/1674-4527/acdfa5",
    journal = "Res. Astron. Astrophys.",
    volume = "23",
    number = "7",
    pages = "075024",
    year = "2023"
}

@article{Smits:2008cf,
    author = "Smits, R. and Kramer, M. and Stappers, B. and Lorimer, D. R. and Cordes, J. and Faulkner, A.",
    title = "{Pulsar searches and timing with the square kilometre array}",
    eprint = "0811.0211",
    archivePrefix = "arXiv",
    primaryClass = "astro-ph",
    doi = "10.1051/0004-6361:200810383",
    journal = "Astron. Astrophys.",
    volume = "493",
    pages = "1161--1170",
    year = "2009"
}

@article{Sesana:2008mz,
    author = "Sesana, Alberto and Vecchio, Alberto and Colacino, Carlo Nicola",
    title = "{The stochastic gravitational-wave background from massive black hole binary systems: implications for observations with Pulsar Timing Arrays}",
    eprint = "0804.4476",
    archivePrefix = "arXiv",
    primaryClass = "astro-ph",
    doi = "10.1111/j.1365-2966.2008.13682.x",
    journal = "Mon. Not. Roy. Astron. Soc.",
    volume = "390",
    pages = "192",
    year = "2008"
}

@article{Spallicci:2011nr,
    author = "Spallicci, Alessandro D. A. M.",
    title = "{On the complementarity of pulsar timing and space laser interferometry for the individual detection of supermassive black hole binaries}",
    eprint = "1107.5984",
    archivePrefix = "arXiv",
    primaryClass = "gr-qc",
    doi = "10.1088/0004-637X/764/2/187",
    journal = "Astrophys. J.",
    volume = "764",
    pages = "187",
    year = "2013"
}

@article{Ellis:2023owy,
    author = {Ellis, John and Fairbairn, Malcolm and H\"utsi, Gert and Raidal, Martti and Urrutia, Juan and Vaskonen, Ville and Veerm\"ae, Hardi},
    title = "{Prospects for future binary black hole gravitational wave studies in light of PTA measurements}",
    eprint = "2301.13854",
    archivePrefix = "arXiv",
    primaryClass = "astro-ph.CO",
    reportNumber = "KCL-PH-TH/2023-04, CERN-TH-2023-008, AION-REPORT/2023-1",
    doi = "10.1051/0004-6361/202346268",
    journal = "Astron. Astrophys.",
    volume = "676",
    pages = "A38",
    year = "2023"
}

@article{LISA:2024hlh,
    author = "Colpi, Monica and others",
    collaboration = "LISA",
    title = "{LISA Definition Study Report}",
    eprint = "2402.07571",
    archivePrefix = "arXiv",
    primaryClass = "astro-ph.CO",
    month = "2",
    year = "2024"
}

@article{Berti:2015itd,
    author = "Berti, Emanuele and others",
    title = "{Testing General Relativity with Present and Future Astrophysical Observations}",
    eprint = "1501.07274",
    archivePrefix = "arXiv",
    primaryClass = "gr-qc",
    doi = "10.1088/0264-9381/32/24/243001",
    journal = "Class. Quant. Grav.",
    volume = "32",
    pages = "243001",
    year = "2015"
}

@article{Yunes:2013dva,
    author = "Yunes, Nicol{\'a}s and Siemens, Xavier",
    title = "{Gravitational-Wave Tests of General Relativity with Ground-Based Detectors and Pulsar Timing-Arrays}",
    eprint = "1304.3473",
    archivePrefix = "arXiv",
    primaryClass = "gr-qc",
    doi = "10.12942/lrr-2013-9",
    journal = "Living Rev. Rel.",
    volume = "16",
    pages = "9",
    year = "2013"
}

@article{LISACosmologyWorkingGroup:2019mwx,
    author = "Belgacem, Enis and others",
    collaboration = "LISA Cosmology Working Group",
    title = "{Testing modified gravity at cosmological distances with LISA standard sirens}",
    eprint = "1906.01593",
    archivePrefix = "arXiv",
    primaryClass = "astro-ph.CO",
    reportNumber = "LISA CosWG-19-01; IFT-UAM-CSIC-19-79, LISA CosWG-19-01",
    doi = "10.1088/1475-7516/2019/07/024",
    journal = "JCAP",
    volume = "07",
    pages = "024",
    year = "2019"
}

@article{Holz:2005df,
    author = "Holz, Daniel E. and Hughes, Scott A.",
    title = "{Using gravitational-wave standard sirens}",
    eprint = "astro-ph/0504616",
    archivePrefix = "arXiv",
    doi = "10.1086/431341",
    journal = "Astrophys. J.",
    volume = "629",
    pages = "15--22",
    year = "2005"
}

@article{Sesana:2010wy,
    author = "Sesana, Alberto and Gair, Jonathan and Berti, Emanuele and Volonteri, Marta",
    title = "{Reconstructing the massive black hole cosmic history through gravitational waves}",
    eprint = "1011.5893",
    archivePrefix = "arXiv",
    primaryClass = "astro-ph.CO",
    doi = "10.1103/PhysRevD.83.044036",
    journal = "Phys. Rev. D",
    volume = "83",
    pages = "044036",
    year = "2011"
}

@article{Klein:2015hvg,
    author = "Klein, Antoine and others",
    title = "{Science with the space-based interferometer eLISA: Supermassive black hole binaries}",
    eprint = "1511.05581",
    archivePrefix = "arXiv",
    primaryClass = "gr-qc",
    doi = "10.1103/PhysRevD.93.024003",
    journal = "Phys. Rev. D",
    volume = "93",
    number = "2",
    pages = "024003",
    year = "2016"
}

@article{Katz:2019qlu,
    author = "Katz, Michael L. and Kelley, Luke Zoltan and Dosopoulou, Fani and Berry, Samantha and Blecha, Laura and Larson, Shane L.",
    title = "{Probing Massive Black Hole Binary Populations with LISA}",
    eprint = "1908.05779",
    archivePrefix = "arXiv",
    primaryClass = "astro-ph.HE",
    doi = "10.1093/mnras/stz3102",
    journal = "Mon. Not. Roy. Astron. Soc.",
    volume = "491",
    number = "2",
    pages = "2301--2317",
    year = "2020"
}

@article{Barausse:2020mdt,
    author = "Barausse, Enrico and Dvorkin, Irina and Tremmel, Michael and Volonteri, Marta and Bonetti, Matteo",
    title = "{Massive Black Hole Merger Rates: The Effect of Kiloparsec Separation Wandering and Supernova Feedback}",
    eprint = "2006.03065",
    archivePrefix = "arXiv",
    primaryClass = "astro-ph.GA",
    doi = "10.3847/1538-4357/abba7f",
    journal = "Astrophys. J.",
    volume = "904",
    number = "1",
    pages = "16",
    year = "2020"
}

@article{Ueda:2014tma,
    author = {Ueda, Yoshihiro and Akiyama, Masayuki and Hasinger, G\"unther and Miyaji, Takamitsu and Watson, Michael G.},
    title = "{Toward the Standard Population Synthesis Model of the X-Ray Background: Evolution of X-Ray Luminosity and Absorption Functions of Active Galactic Nuclei Including Compton-Thick Populations}",
    eprint = "1402.1836",
    archivePrefix = "arXiv",
    primaryClass = "astro-ph.CO",
    doi = "10.1088/0004-637X/786/2/104",
    journal = "Astrophys. J.",
    volume = "786",
    pages = "104",
    year = "2014"
}

@ARTICLE{2011ApJ...737..101L,
       author = {{Liu}, Xin and {Shen}, Yue and {Strauss}, Michael A. and {Hao}, Lei},
        title = "{Active Galactic Nucleus Pairs from the Sloan Digital Sky Survey. I. The Frequency on \raisebox{-0.5ex}\textasciitilde5-100 kpc Scales}",
      journal = {\apj},
         year = 2011,
        month = aug,
       volume = {737},
       number = {2},
          eid = {101},
        pages = {101},
          doi = {10.1088/0004-637X/737/2/101},
archivePrefix = {arXiv},
       eprint = {1104.0950},
 primaryClass = {astro-ph.CO},
       adsurl = {https://ui.adsabs.harvard.edu/abs/2011ApJ...737..101L}
}

@ARTICLE{2012ApJ...746L..22K,
       author = {{Koss}, Michael and {Mushotzky}, Richard and {Treister}, Ezequiel and {Veilleux}, Sylvain and {Vasudevan}, Ranjan and {Trippe}, Margaret},
        title = "{Understanding Dual Active Galactic Nucleus Activation in the nearby Universe}",
      journal = {\apjl},
         year = 2012,
        month = feb,
       volume = {746},
       number = {2},
          eid = {L22},
        pages = {L22},
          doi = {10.1088/2041-8205/746/2/L22},
archivePrefix = {arXiv},
       eprint = {1201.2944},
 primaryClass = {astro-ph.HE},
       adsurl = {https://ui.adsabs.harvard.edu/abs/2012ApJ...746L..22K}
}

@ARTICLE{2020ApJ...899..154S,
       author = "Silverman, John D. and others",
        title = "{Dual Supermassive Black Holes at Close Separation Revealed by the Hyper Suprime-Cam Subaru Strategic Program}",
      journal = {\apj},
         year = 2020,
        month = aug,
       volume = {899},
       number = {2},
          eid = {154},
        pages = {154},
          doi = {10.3847/1538-4357/aba4a3},
archivePrefix = {arXiv},
       eprint = {2007.05581},
 primaryClass = {astro-ph.GA},
       adsurl = {https://ui.adsabs.harvard.edu/abs/2020ApJ...899..154S}
}

@article{Shen:2022cmp,
    author = "Shen, Yue and others",
    title = "{Statistics of Galactic-scale Quasar Pairs at Cosmic Noon}",
    eprint = "2208.04979",
    archivePrefix = "arXiv",
    primaryClass = "astro-ph.GA",
    doi = "10.3847/1538-4357/aca662",
    journal = "Astrophys. J.",
    volume = "943",
    number = "1",
    pages = "38",
    year = "2023"
}

@ARTICLE{2023arXiv231003067P,
       author = "Perna, Michele and others",
      journal = {arXiv e-prints},
         year = 2023,
        month = oct,
          eid = {arXiv:2310.03067},
        pages = {arXiv:2310.03067},
          doi = {10.48550/arXiv.2310.03067},
archivePrefix = {arXiv},
       eprint = {2310.03067},
 primaryClass = {astro-ph.GA},
       adsurl = {https://ui.adsabs.harvard.edu/abs/2023arXiv231003067P}
}

@ARTICLE{2024arXiv240514980L,
       author = {{Li}, Junyao and {Zhuang}, Ming-Yang and {Shen}, Yue and {Volonteri}, Marta and {Chen}, Nianyi and {Di Matteo}, Tiziana},
        title = "{Active Galactic Nuclei and Host Galaxies in COSMOS-Web. II. First Look at the Kpc-scale Dual and Offset AGN Population}",
      journal = {arXiv e-prints},
         year = 2024,
        month = may,
          eid = {arXiv:2405.14980},
        pages = {arXiv:2405.14980},
          doi = {10.48550/arXiv.2405.14980},
archivePrefix = {arXiv},
       eprint = {2405.14980},
 primaryClass = {astro-ph.GA},
       adsurl = {https://ui.adsabs.harvard.edu/abs/2024arXiv240514980L}
}

@article{Chen:2023hyj,
    author = "Chen, Yunfeng and Yu, Qingjuan and Lu, Youjun",
    title = "{Pulsar Timing Array Detections of Supermassive Binary Black Holes: Implications from the Detected Common Process Signal and Beyond}",
    eprint = "2306.10997",
    archivePrefix = "arXiv",
    primaryClass = "astro-ph.HE",
    doi = "10.3847/1538-4357/ace59f",
    journal = "Astrophys. J.",
    volume = "955",
    number = "2",
    pages = "132",
    year = "2023"
}

@article{Chen:2023zkb,
    author = "Chen, Zu-Cheng and Huang, Qing-Guo and Liu, Chang and Liu, Lang and Liu, Xiao-Jin and Wu, You and Wu, Yu-Mei and Yi, Zhu and You, Zhi-Qiang",
    title = "{Prospects for Taiji to detect a gravitational-wave background from cosmic strings}",
    eprint = "2310.00411",
    archivePrefix = "arXiv",
    primaryClass = "astro-ph.IM",
    doi = "10.1088/1475-7516/2024/03/022",
    journal = "JCAP",
    volume = "03",
    pages = "022",
    year = "2024"
}

@article{TianQin:2015yph,
    author = "Luo, Jun and others",
    collaboration = "TianQin",
    title = "{TianQin: a space-borne gravitational wave detector}",
    eprint = "1512.02076",
    archivePrefix = "arXiv",
    primaryClass = "astro-ph.IM",
    doi = "10.1088/0264-9381/33/3/035010",
    journal = "Class. Quant. Grav.",
    volume = "33",
    number = "3",
    pages = "035010",
    year = "2016"
}

@article{Gong:2021gvw,
    author = "Gong, Yungui and Luo, Jun and Wang, Bin",
    title = "{Concepts and status of Chinese space gravitational wave detection projects}",
    eprint = "2109.07442",
    archivePrefix = "arXiv",
    primaryClass = "astro-ph.IM",
    doi = "10.1038/s41550-021-01480-3",
    journal = "Nature Astron.",
    volume = "5",
    number = "9",
    pages = "881--889",
    year = "2021"
}

@article{Liang:2019pry,
    author = "Liang, Dicong and Gong, Yungui and Weinstein, Alan J. and Zhang, Chao and Zhang, Chunyu",
    title = "{Frequency response of space-based interferometric gravitational-wave detectors}",
    eprint = "1901.09624",
    archivePrefix = "arXiv",
    primaryClass = "gr-qc",
    doi = "10.1103/PhysRevD.99.104027",
    journal = "Phys. Rev. D",
    volume = "99",
    number = "10",
    pages = "104027",
    year = "2019"
}

@article{Lange:2014kca,
    author = "Lange, Rebecca and others",
    title = "{Galaxy And Mass Assembly (GAMA): mass-size relations of z$<$0.1 galaxies subdivided by S{\'e}rsic index, colour and morphology}",
    eprint = "1411.6355",
    archivePrefix = "arXiv",
    primaryClass = "astro-ph.GA",
    doi = "10.1093/mnras/stu2467",
    journal = "Mon. Not. Roy. Astron. Soc.",
    volume = "447",
    pages = "2603",
    year = "2015"
}

@article{Zhang:2020khm,
    author = "Zhang, Chunyu and Gao, Qing and Gong, Yungui and Wang, Bin and Weinstein, Alan J. and Zhang, Chao",
    title = "{Full analytical formulas for frequency response of space-based gravitational wave detectors}",
    eprint = "2003.01441",
    archivePrefix = "arXiv",
    primaryClass = "gr-qc",
    doi = "10.1103/PhysRevD.101.124027",
    journal = "Phys. Rev. D",
    volume = "101",
    number = "12",
    pages = "124027",
    year = "2020"
}

@article{Furusawa:2023fwl,
    author = "Furusawa, Kazuya and Tashiro, Hiroyuki and Yokoyama, Shuichiro and Ichiki, Kiyotomo",
    title = "{Probing the Mass Relation between Supermassive Black Holes and Dark Matter Halos at High Redshifts by Gravitational Wave Experiments}",
    eprint = "2306.16281",
    archivePrefix = "arXiv",
    primaryClass = "astro-ph.CO",
    doi = "10.3847/1538-4357/ad088f",
    journal = "Astrophys. J.",
    volume = "959",
    number = "2",
    pages = "117",
    year = "2023"
}

@article{Sato-Polito:2024lew,
    author = "Sato-Polito, Gabriela and Zaldarriaga, Matias",
    title = "{The distribution of the gravitational-wave background from supermassive black holes}",
    eprint = "2406.17010",
    archivePrefix = "arXiv",
    primaryClass = "astro-ph.CO",
    month = "6",
    year = "2024"
}

@article{Kaiser:2020tlg,
    author = "Kaiser, Andrew R. and McWilliams, Sean T.",
    title = "{Sensitivity of present and future detectors across the black-hole binary gravitational wave spectrum}",
    eprint = "2010.02135",
    archivePrefix = "arXiv",
    primaryClass = "gr-qc",
    doi = "10.1088/1361-6382/abd4f6",
    journal = "Class. Quant. Grav.",
    volume = "38",
    number = "5",
    pages = "055009",
    year = "2021"
}

@article{Amaro-Seoane:2012vvq,
    author = "Amaro-Seoane, Pau and others",
    editor = "Hannam, Mark and Sutton, Patrick and Hild, Stefan and van den Broeck, Chris",
    title = "{Low-frequency gravitational-wave science with eLISA/NGO}",
    eprint = "1202.0839",
    archivePrefix = "arXiv",
    primaryClass = "gr-qc",
    doi = "10.1088/0264-9381/29/12/124016",
    journal = "Class. Quant. Grav.",
    volume = "29",
    pages = "124016",
    year = "2012"
}

@article{DOrazio:2017dyb,
    author = "D'Orazio, Daniel J. and Loeb, Abraham",
    title = "{Repeated Imaging of Massive Black Hole Binary Orbits with Millimeter Interferometry: Measuring Black Hole Masses and the Hubble Constant}",
    eprint = "1712.02362",
    archivePrefix = "arXiv",
    primaryClass = "astro-ph.HE",
    doi = "10.3847/1538-4357/aad413",
    journal = "Astrophys. J.",
    volume = "863",
    number = "2",
    pages = "185",
    year = "2018"
}

@article{Zhao:2023kff,
    author = "Zhao, Shan-Shan and Jiang, Wu and Lu, Ru-Sen and Huang, Lei and Shen, Zhi-Qiang",
    title = "{How Many Supermassive Black Hole Binaries Are Detectable through Tracking Relative Motions by (Sub)millimeter Very Long Baseline Interferometry?}",
    eprint = "2311.11589",
    archivePrefix = "arXiv",
    primaryClass = "astro-ph.GA",
    doi = "10.3847/1538-4357/ad0da1",
    journal = "Astrophys. J.",
    volume = "961",
    number = "1",
    pages = "20",
    year = "2024"
}

@article{Guo:2024tlg,
    author = "Guo, Xiao and Yu, Qingjuan and Lu, Youjun",
    title = "{Constraining the Binarity of Massive Black Holes in the Galactic Center and Some Nearby Galaxies via Pulsar Timing Array Observations of Gravitational Waves}",
    eprint = "2411.14150",
    archivePrefix = "arXiv",
    primaryClass = "astro-ph.HE",
    month = "11",
    year = "2024"
}

@article{Robson:2018ifk,
    author = "Robson, Travis and Cornish, Neil J. and Liu, Chang",
    title = "{The construction and use of LISA sensitivity curves}",
    eprint = "1803.01944",
    archivePrefix = "arXiv",
    primaryClass = "astro-ph.HE",
    doi = "10.1088/1361-6382/ab1101",
    journal = "Class. Quant. Grav.",
    volume = "36",
    number = "10",
    pages = "105011",
    year = "2019"
}

@article{Arun:2008zn,
    author = "Arun, K. G. and others",
    editor = "Lobo, Alberto and Sopuerta, Carlos F.",
    title = "{Massive Black Hole Binary Inspirals: Results from the LISA Parameter Estimation Taskforce}",
    eprint = "0811.1011",
    archivePrefix = "arXiv",
    primaryClass = "gr-qc",
    doi = "10.1088/0264-9381/26/9/094027",
    journal = "Class. Quant. Grav.",
    volume = "26",
    pages = "094027",
    year = "2009"
}

@article{Janssen:2014dka,
    author = "Janssen, Gemma and others",
    editor = "Bourke, Tyler L. and others",
    title = "{Gravitational wave astronomy with the SKA}",
    eprint = "1501.00127",
    archivePrefix = "arXiv",
    primaryClass = "astro-ph.IM",
    doi = "10.22323/1.215.0037",
    journal = "PoS",
    volume = "AASKA14",
    pages = "037",
    year = "2015"
}

@article{Babak:2024yhu,
    author = "Babak, Stanislav and Falxa, Mikel and Franciolini, Gabriele and Pieroni, Mauro",
    title = "{Forecasting the sensitivity of pulsar timing arrays to gravitational wave backgrounds}",
    eprint = "2404.02864",
    archivePrefix = "arXiv",
    primaryClass = "astro-ph.CO",
    reportNumber = "CERN-TH-2024-039",
    doi = "10.1103/PhysRevD.110.063022",
    journal = "Phys. Rev. D",
    volume = "110",
    number = "6",
    pages = "063022",
    year = "2024"
}

@ARTICLE{2017NatAs...1..886M,
       author = {{Mingarelli}, Chiara M.~F. and {Lazio}, T. Joseph W. and {Sesana}, Alberto and {Greene}, Jenny E. and {Ellis}, Justin A. and {Ma}, Chung-Pei and {Croft}, Steve and {Burke-Spolaor}, Sarah and {Taylor}, Stephen R.},
        title = "{The local nanohertz gravitational-wave landscape from supermassive black hole binaries}",
      journal = {Nature Astronomy},
         year = 2017,
        month = nov,
       volume = {1},
        pages = {886-892},
          doi = {10.1038/s41550-017-0299-6},
archivePrefix = {arXiv},
       eprint = {1708.03491},
 primaryClass = {astro-ph.GA},
       adsurl = {https://ui.adsabs.harvard.edu/abs/2017NatAs...1..886M}
}

@article{Bell:2003cj,
    author = "Bell, Eric F. and McIntosh, Daniel H. and Katz, Neal and Weinberg, Martin D.",
    title = "{The optical and near-infrared properties of galaxies. 1. Luminosity and stellar mass functions}",
    eprint = "astro-ph/0302543",
    archivePrefix = "arXiv",
    doi = "10.1086/378847",
    journal = "Astrophys. J. Suppl.",
    volume = "149",
    pages = "289",
    year = "2003"
}

@article{Muzzin:2013yxa,
    author = "Muzzin, Adam and others",
    title = "{The Evolution of the Stellar Mass Functions of Star-Forming and Quiescent Galaxies to z = 4 from the COSMOS/UltraVISTA Survey}",
    eprint = "1303.4409",
    archivePrefix = "arXiv",
    primaryClass = "astro-ph.CO",
    doi = "10.1088/0004-637X/777/1/18",
    journal = "Astrophys. J.",
    volume = "777",
    pages = "18",
    year = "2013"
}

@ARTICLE{2019MNRAS.488.3143B,
       author = {{Behroozi}, Peter and {Wechsler}, Risa H. and {Hearin}, Andrew P. and {Conroy}, Charlie},
        title = "{UNIVERSEMACHINE: The correlation between galaxy growth and dark matter halo assembly from z = 0-10}",
      journal = {\mnras},
         year = 2019,
        month = sep,
       volume = {488},
       number = {3},
        pages = {3143-3194},
          doi = {10.1093/mnras/stz1182},
archivePrefix = {arXiv},
       eprint = {1806.07893},
 primaryClass = {astro-ph.GA},
       adsurl = {https://ui.adsabs.harvard.edu/abs/2019MNRAS.488.3143B}
}

@article{Rodriguez-Gomez:2015aua,
    author = "Rodriguez-Gomez, Vicente and others",
    title = "{The merger rate of galaxies in the Illustris Simulation: a comparison with observations and semi-empirical models}",
    eprint = "1502.01339",
    archivePrefix = "arXiv",
    primaryClass = "astro-ph.GA",
    doi = "10.1093/mnras/stv264",
    journal = "Mon. Not. Roy. Astron. Soc.",
    volume = "449",
    number = "1",
    pages = "49--64",
    year = "2015"
}

@article{LSST:2008ijt,
    author = "Ivezi\'c, \v{Z}eljko and others",
    collaboration = "LSST",
    title = "{LSST: from Science Drivers to Reference Design and Anticipated Data Products}",
    eprint = "0805.2366",
    archivePrefix = "arXiv",
    primaryClass = "astro-ph",
    reportNumber = "SLAC-PUB-16076",
    doi = "10.3847/1538-4357/ab042c",
    journal = "Astrophys. J.",
    volume = "873",
    number = "2",
    pages = "111",
    year = "2019"
}

@article{TMTInternationalScienceDevelopmentTeamsTMTScienceAdvisoryCommittee:2015pvw,
    author = "Skidmore, Warren and others",
    collaboration = "TMT International Science Development Teams \& TMT Science Advisory Committee",
    title = "{Thirty Meter Telescope Detailed Science Case: 2015}",
    eprint = "1505.01195",
    archivePrefix = "arXiv",
    primaryClass = "astro-ph.IM",
    reportNumber = "TMT.PSC.TEC.07.007.REL02, TMT.PSC.TEC.07.007.CCR04",
    doi = "10.1088/1674-4527/15/12/001",
    journal = "Res. Astron. Astrophys.",
    volume = "15",
    number = "12",
    pages = "1945--2140",
    year = "2015"
}

@INPROCEEDINGS{2012SPIE.8444E..1HJ,
       author = "Johns, Matt and others",
        title = "{Giant Magellan Telescope: overview}",
    booktitle = {Ground-based and Airborne Telescopes IV},
         year = 2012,
       editor = {{Stepp}, Larry M. and {Gilmozzi}, Roberto and {Hall}, Helen J.},
       series = {Society of Photo-Optical Instrumentation Engineers (SPIE) Conference Series},
       volume = {8444},
        month = sep,
          eid = {84441H},
        pages = {84441H},
          doi = {10.1117/12.926716},
       adsurl = {https://ui.adsabs.harvard.edu/abs/2012SPIE.8444E..1HJ}
}

@article{Padovani:2023dxc,
    author = "Padovani, Paolo and Cirasuolo, Michele",
    title = "{The Extremely Large Telescope}",
    eprint = "2312.04299",
    archivePrefix = "arXiv",
    primaryClass = "astro-ph.IM",
    doi = "10.1080/00107514.2023.2266921",
    journal = "Contemp. Phys.",
    volume = "64",
    number = "1",
    pages = "47--64",
    year = "2023"
}

@ARTICLE{2019astro2020T.504K,
       author = {{Koss}, Michael and {U}, Vivian and {Hodges-Kluck}, Edmund and {Blecha}, Laura and {Kartaltepe}, Jeyhan and {Kocevski}, Dale and {Comerford}, Julia M. and {Barrows}, R. Scott and {Cicone}, Claudia and {Muller-Sanchez}, Francisco and {Treister}, Ezequiel and {Gultekin}, Kayhan and {Ricci}, Claudio and {Foord}, Adi and {Lotz}, Jennifer},
        title = "{Black Hole Growth in Mergers and Dual AGN}",
      journal = {Astro2020: Decadal Survey on Astronomy and Astrophysics},
         year = 2019,
        month = may,
       volume = {2020},
        pages = {504},
          doi = {10.48550/arXiv.1903.06720},
archivePrefix = {arXiv},
       eprint = {1903.06720},
 primaryClass = {astro-ph.GA},
       adsurl = {https://ui.adsabs.harvard.edu/abs/2019astro2020T.504K}
}

@article{Sesana:2013wja,
    author = "Sesana, A.",
    title = "{Insights into the astrophysics of supermassive black hole binaries from pulsar timing observations}",
    eprint = "1307.2600",
    archivePrefix = "arXiv",
    primaryClass = "astro-ph.CO",
    doi = "10.1088/0264-9381/30/22/224014",
    journal = "Class. Quant. Grav.",
    volume = "30",
    pages = "224014",
    year = "2013"
}

@article{Larson:1999we,
    author = "Larson, Shane L. and Hiscock, William A. and Hellings, Ronald W.",
    title = "{Sensitivity curves for spaceborne gravitational wave interferometers}",
    eprint = "gr-qc/9909080",
    archivePrefix = "arXiv",
    reportNumber = "MSUPHY-99-04",
    doi = "10.1103/PhysRevD.62.062001",
    journal = "Phys. Rev. D",
    volume = "62",
    pages = "062001",
    year = "2000"
}

@ARTICLE{2019ApJ...872..104N,
       author = {{Nguyen}, Dieu D. and {Seth}, Anil C. and {Neumayer}, Nadine and {Iguchi}, Satoru and {Cappellari}, Michelle and {Strader}, Jay and {Chomiuk}, Laura and {Tremou}, Evangelia and {Pacucci}, Fabio and {Nakanishi}, Kouichiro and {Bahramian}, Arash and {Nguyen}, Phuong M. and {den Brok}, Mark and {Ahn}, Christopher C. and {Voggel}, Karina T. and {Kacharov}, Nikolay and {Tsukui}, Takafumi and {Ly}, Cuc K. and {Dumont}, Antoine and {Pechetti}, Renuka},
        title = "{Improved Dynamical Constraints on the Masses of the Central Black Holes in Nearby Low-mass Early-type Galactic Nuclei and the First Black Hole Determination for NGC 205}",
      journal = {\apj},
         year = 2019,
        month = feb,
       volume = {872},
       number = {1},
          eid = {104},
        pages = {104},
          doi = {10.3847/1538-4357/aafe7a},
archivePrefix = {arXiv},
       eprint = {1901.05496},
 primaryClass = {astro-ph.GA},
       adsurl = {https://ui.adsabs.harvard.edu/abs/2019ApJ...872..104N}
}

@ARTICLE{2017arXiv170200786A,
       author = {{Amaro-Seoane}, Pau and {Audley}, Heather and {Babak}, Stanislav and {Baker}, John and {Barausse}, Enrico and {Bender}, Peter and {Berti}, Emanuele and {Binetruy}, Pierre and {Born}, Michael and {Bortoluzzi}, Daniele and {Camp}, Jordan and {Caprini}, Chiara and {Cardoso}, Vitor and {Colpi}, Monica and {Conklin}, John and {Cornish}, Neil and {Cutler}, Curt and {Danzmann}, Karsten and {Dolesi}, Rita and {Ferraioli}, Luigi and {Ferroni}, Valerio and {Fitzsimons}, Ewan and {Gair}, Jonathan and {Gesa Bote}, Lluis and {Giardini}, Domenico and {Gibert}, Ferran and {Grimani}, Catia and {Halloin}, Hubert and {Heinzel}, Gerhard and {Hertog}, Thomas and {Hewitson}, Martin and {Holley-Bockelmann}, Kelly and {Hollington}, Daniel and {Hueller}, Mauro and {Inchauspe}, Henri and {Jetzer}, Philippe and {Karnesis}, Nikos and {Killow}, Christian and {Klein}, Antoine and {Klipstein}, Bill and {Korsakova}, Natalia and {Larson}, Shane L and {Livas}, Jeffrey and {Lloro}, Ivan and {Man}, Nary and {Mance}, Davor and {Martino}, Joseph and {Mateos}, Ignacio and {McKenzie}, Kirk and {McWilliams}, Sean T and {Miller}, Cole and {Mueller}, Guido and {Nardini}, Germano and {Nelemans}, Gijs and {Nofrarias}, Miquel and {Petiteau}, Antoine and {Pivato}, Paolo and {Plagnol}, Eric and {Porter}, Ed and {Reiche}, Jens and {Robertson}, David and {Robertson}, Norna and {Rossi}, Elena and {Russano}, Giuliana and {Schutz}, Bernard and {Sesana}, Alberto and {Shoemaker}, David and {Slutsky}, Jacob and {Sopuerta}, Carlos F. and {Sumner}, Tim and {Tamanini}, Nicola and {Thorpe}, Ira and {Troebs}, Michael and {Vallisneri}, Michele and {Vecchio}, Alberto and {Vetrugno}, Daniele and {Vitale}, Stefano and {Volonteri}, Marta and {Wanner}, Gudrun and {Ward}, Harry and {Wass}, Peter and {Weber}, William and {Ziemer}, John and {Zweifel}, Peter},
        title = "{Laser Interferometer Space Antenna}",
      journal = {arXiv e-prints},
         year = 2017,
        month = feb,
          eid = {arXiv:1702.00786},
        pages = {arXiv:1702.00786},
          doi = {10.48550/arXiv.1702.00786},
archivePrefix = {arXiv},
       eprint = {1702.00786},
 primaryClass = {astro-ph.IM},
       adsurl = {https://ui.adsabs.harvard.edu/abs/2017arXiv170200786A}
}

@article{Genel:2014lma,
    author = "Genel, Shy and Vogelsberger, Mark and Springel, Volker and Sijacki, Debora and Nelson, Dylan and Snyder, Greg and Rodriguez-Gomez, Vicente and Torrey, Paul and Hernquist, Lars",
    title = "{Introducing the Illustris Project: the evolution of galaxy populations across cosmic time}",
    eprint = "1405.3749",
    archivePrefix = "arXiv",
    primaryClass = "astro-ph.CO",
    doi = "10.1093/mnras/stu1654",
    journal = "Mon. Not. Roy. Astron. Soc.",
    volume = "445",
    number = "1",
    pages = "175--200",
    year = "2014"
}

@article{Kormendy:2013dxa,
    author = "Kormendy, John and Ho, Luis C.",
    title = "{Coevolution (Or Not) of Supermassive Black Holes and Host Galaxies}",
    eprint = "1304.7762",
    archivePrefix = "arXiv",
    primaryClass = "astro-ph.CO",
    doi = "10.1146/annurev-astro-082708-101811",
    journal = "Ann. Rev. Astron. Astrophys.",
    volume = "51",
    pages = "511--653",
    year = "2013"
}

@article{Porayko:2018sfa,
    author = "Porayko, Nataliya K. and others",
    title = "{Parkes Pulsar Timing Array constraints on ultralight scalar-field dark matter}",
    eprint = "1810.03227",
    archivePrefix = "arXiv",
    primaryClass = "astro-ph.CO",
    doi = "10.1103/PhysRevD.98.102002",
    journal = "Phys. Rev. D",
    volume = "98",
    number = "10",
    pages = "102002",
    year = "2018"
}

@article{Cai:2023ywp,
    author = "Cai, Rong-Gen and Guo, Zong-Kuan and Hu, Bin and Liu, Chang and Lu, Youjun and Ni, Wei-Tou and Ruan, Wen-Hong and Seto, Naoki and Wang, Gang and Wu, Yue-Liang",
    title = "{On networks of space-based gravitational-wave detectors}",
    eprint = "2305.04551",
    archivePrefix = "arXiv",
    primaryClass = "gr-qc",
    doi = "10.1016/j.fmre.2023.10.007",
    journal = "Fund. Res.",
    volume = "4",
    pages = "1072--1085",
    year = "2024"
}

@article{Ruan:2019tje,
    author = "Ruan, Wen-Hong and Liu, Chang and Guo, Zong-Kuan and Wu, Yue-Liang and Cai, Rong-Gen",
    title = "{The LISA-Taiji Network: Precision Localization of Coalescing Massive Black Hole Binaries}",
    eprint = "1909.07104",
    archivePrefix = "arXiv",
    primaryClass = "gr-qc",
    doi = "10.34133/2021/6014164",
    journal = "Research",
    volume = "2021",
    pages = "6014164",
    year = "2021"
}

@article{Ruan:2020smc,
    author = "Ruan, Wen-Hong and Liu, Chang and Guo, Zong-Kuan and Wu, Yue-Liang and Cai, Rong-Gen",
    title = "{The LISA-Taiji network}",
    eprint = "2002.03603",
    archivePrefix = "arXiv",
    primaryClass = "gr-qc",
    doi = "10.1038/s41550-019-1008-4",
    journal = "Nature Astron.",
    volume = "4",
    pages = "108--109",
    year = "2020"
}

@article{Vilenkin:1984ib,
    author = "Vilenkin, Alexander",
    title = "{Cosmic Strings and Domain Walls}",
    reportNumber = "PRINT-84-0840 (TUFTS)",
    doi = "10.1016/0370-1573(85)90033-X",
    journal = "Phys. Rept.",
    volume = "121",
    pages = "263--315",
    year = "1985"
}

@article{Hindmarsh:1994re,
    author = "Hindmarsh, M. B. and Kibble, T. W. B.",
    title = "{Cosmic strings}",
    eprint = "hep-ph/9411342",
    archivePrefix = "arXiv",
    reportNumber = "SUSX-TP-94-74, IMPERIAL-TP-94-95-5, NI-94025",
    doi = "10.1088/0034-4885/58/5/001",
    journal = "Rept. Prog. Phys.",
    volume = "58",
    pages = "477--562",
    year = "1995"
}

@article{Caprini:2015zlo,
    author = "Caprini, Chiara and others",
    title = "{Science with the space-based interferometer eLISA. II: Gravitational waves from cosmological phase transitions}",
    eprint = "1512.06239",
    archivePrefix = "arXiv",
    primaryClass = "astro-ph.CO",
    reportNumber = "DESY-15-246",
    doi = "10.1088/1475-7516/2016/04/001",
    journal = "JCAP",
    volume = "04",
    pages = "001",
    year = "2016"
}

@article{Saikawa:2017hiv,
    author = "Saikawa, Ken'ichi",
    title = "{A review of gravitational waves from cosmic domain walls}",
    eprint = "1703.02576",
    archivePrefix = "arXiv",
    primaryClass = "hep-ph",
    reportNumber = "DESY-17-036",
    doi = "10.3390/universe3020040",
    journal = "Universe",
    volume = "3",
    number = "2",
    pages = "40",
    year = "2017"
}

@article{Caprini:2019egz,
    author = "Caprini, Chiara and others",
    title = "{Detecting gravitational waves from cosmological phase transitions with LISA: an update}",
    eprint = "1910.13125",
    archivePrefix = "arXiv",
    primaryClass = "astro-ph.CO",
    reportNumber = "DESY-19-159, IPPP/19/27, HIP-2019-14/TH, MITP/19-066, IFT-UAM/CSIC-19-139",
    doi = "10.1088/1475-7516/2020/03/024",
    journal = "JCAP",
    volume = "03",
    pages = "024",
    year = "2020"
}

@article{Hindmarsh:2020hop,
    author = {Hindmarsh, Mark B. and L\"uben, Marvin and Lumma, Johannes and Pauly, Martin},
    title = "{Phase transitions in the early universe}",
    eprint = "2008.09136",
    archivePrefix = "arXiv",
    primaryClass = "astro-ph.CO",
    reportNumber = "MPP-2020-163, HIP-2020-27/TH",
    doi = "10.21468/SciPostPhysLectNotes.24",
    journal = "SciPost Phys. Lect. Notes",
    volume = "24",
    pages = "1",
    year = "2021"
}

@article{Domenech:2021ztg,
    author = "Dom\`enech, Guillem",
    title = "{Scalar Induced Gravitational Waves Review}",
    eprint = "2109.01398",
    archivePrefix = "arXiv",
    primaryClass = "gr-qc",
    doi = "10.3390/universe7110398",
    journal = "Universe",
    volume = "7",
    number = "11",
    pages = "398",
    year = "2021"
}

@article{Yuan:2021qgz,
    author = "Yuan, Chen and Huang, Qing-Guo",
    title = "{A topic review on probing primordial black hole dark matter with scalar induced gravitational waves}",
    eprint = "2103.04739",
    archivePrefix = "arXiv",
    primaryClass = "astro-ph.GA",
    doi = "10.1016/j.isci.2021.102860",
    journal = "iScience",
    volume = "24",
    pages = "102860",
    year = "2021"
}

@article{Khan:2015jqa,
    author = {Khan, Sebastian and Husa, Sascha and Hannam, Mark and Ohme, Frank and P\"urrer, Michael and Jim\'enez Forteza, Xisco and Boh\'e, Alejandro},
    title = "{Frequency-domain gravitational waves from nonprecessing black-hole binaries. II. A phenomenological model for the advanced detector era}",
    eprint = "1508.07253",
    archivePrefix = "arXiv",
    primaryClass = "gr-qc",
    doi = "10.1103/PhysRevD.93.044007",
    journal = "Phys. Rev. D",
    volume = "93",
    number = "4",
    pages = "044007",
    year = "2016"
}

@article{Husa:2015iqa,
    author = {Husa, Sascha and Khan, Sebastian and Hannam, Mark and P\"urrer, Michael and Ohme, Frank and Jim\'enez Forteza, Xisco and Boh\'e, Alejandro},
    title = "{Frequency-domain gravitational waves from nonprecessing black-hole binaries. I. New numerical waveforms and anatomy of the signal}",
    eprint = "1508.07250",
    archivePrefix = "arXiv",
    primaryClass = "gr-qc",
    doi = "10.1103/PhysRevD.93.044006",
    journal = "Phys. Rev. D",
    volume = "93",
    number = "4",
    pages = "044006",
    year = "2016"
}

@article{Thrane:2013oya,
    author = "Thrane, Eric and Romano, Joseph D.",
    title = "{Sensitivity curves for searches for gravitational-wave backgrounds}",
    eprint = "1310.5300",
    archivePrefix = "arXiv",
    primaryClass = "astro-ph.IM",
    doi = "10.1103/PhysRevD.88.124032",
    journal = "Phys. Rev. D",
    volume = "88",
    number = "12",
    pages = "124032",
    year = "2013"
}

@article{Lam:2018uta,
    author = "Lam, M. T.",
    title = "{Optimizing Pulsar Timing Array Observational Cadences for Sensitivity to Low-frequency Gravitational-wave Sources}",
    eprint = "1808.10071",
    archivePrefix = "arXiv",
    primaryClass = "astro-ph.HE",
    doi = "10.3847/1538-4357/aae533",
    journal = "Astrophys. J.",
    volume = "868",
    number = "1",
    pages = "33",
    year = "2018"
}

@article{Moore:2014lga,
    author = "Moore, C. J. and Cole, R. H. and Berry, C. P. L.",
    title = "{Gravitational-wave sensitivity curves}",
    eprint = "1408.0740",
    archivePrefix = "arXiv",
    primaryClass = "gr-qc",
    reportNumber = "LIGO-P1400129",
    doi = "10.1088/0264-9381/32/1/015014",
    journal = "Class. Quant. Grav.",
    volume = "32",
    number = "1",
    pages = "015014",
    year = "2015"
}

@article{Hazboun:2019vhv,
    author = "Hazboun, Jeffrey S. and Romano, Joseph D. and Smith, Tristan L.",
    title = "{Realistic sensitivity curves for pulsar timing arrays}",
    eprint = "1907.04341",
    archivePrefix = "arXiv",
    primaryClass = "gr-qc",
    doi = "10.1103/PhysRevD.100.104028",
    journal = "Phys. Rev. D",
    volume = "100",
    number = "10",
    pages = "104028",
    year = "2019"
}

@article{Curylo:2021pvf,
    author = "Cury\l{}o, Ma\l{}gorzata and Bulik, Tomasz",
    title = "{Predictions for LISA and PTA based on SHARK galaxy simulations}",
    eprint = "2108.11232",
    archivePrefix = "arXiv",
    primaryClass = "astro-ph.CO",
    doi = "10.1051/0004-6361/202141987",
    journal = "Astron. Astrophys.",
    volume = "660",
    pages = "A68",
    year = "2022",
    note = "[Erratum: Astron.Astrophys. 686, C4 (2024)]"
}

@article{Casey-Clyde:2021xro,
    author = "Casey-Clyde, J. Andrew and Mingarelli, Chiara M. F. and Greene, Jenny E. and Pardo, Kris and Na\~nez, Morgan and Goulding, Andy D.",
    title = "{A Quasar-based Supermassive Black Hole Binary Population Model: Implications for the Gravitational Wave Background}",
    eprint = "2107.11390",
    archivePrefix = "arXiv",
    primaryClass = "astro-ph.HE",
    doi = "10.3847/1538-4357/ac32de",
    journal = "Astrophys. J.",
    volume = "924",
    number = "2",
    pages = "93",
    year = "2022"
}

@article{Kis-Toth:2024gkm,
    author = "Kis-T\'oth, \'Agnes and Haiman, Zolt\'an and Frei, Zsolt",
    title = "{Can quasars, triggered by mergers, account for NANOGrav's stochastic gravitational wave background?}",
    eprint = "2412.12726",
    archivePrefix = "arXiv",
    primaryClass = "astro-ph.CO",
    month = "12",
    year = "2024"
}

@article{Kusakabe:2025rmn,
    author = "Kusakabe, Katsunori and Inoue, Yoshiyuki and Toyouchi, Daisuke",
    title = "{Coherence of Supermassive Black Hole Binary Demographics with the nHz Stochastic Gravitational Wave Background}",
    eprint = "2510.10548",
    archivePrefix = "arXiv",
    primaryClass = "astro-ph.HE",
    month = "10",
    year = "2025"
}

@article{Planck:2018vyg,
    author = "Aghanim, N. and others",
    collaboration = "Planck",
    title = "{Planck 2018 results. VI. Cosmological parameters}",
    eprint = "1807.06209",
    archivePrefix = "arXiv",
    primaryClass = "astro-ph.CO",
    doi = "10.1051/0004-6361/201833910",
    journal = "Astron. Astrophys.",
    volume = "641",
    pages = "A6",
    year = "2020",
    note = "[Erratum: Astron.Astrophys. 652, C4 (2021)]"
}

@INPROCEEDINGS{2011ASSP...21..229H,
       author = {{Hobbs}, George},
        title = "{Pulsars as gravitational wave detectors}",
    booktitle = {High-Energy Emission from Pulsars and their Systems},
         year = 2011,
       editor = {{Torres}, Diego F. and {Rea}, Nanda},
       series = {Astrophysics and Space Science Proceedings},
       volume = {21},
        month = jan,
        pages = {229},
          doi = {10.1007/978-3-642-17251-9_20},
archivePrefix = {arXiv},
       eprint = {1006.3969},
 primaryClass = {astro-ph.SR},
       adsurl = {https://ui.adsabs.harvard.edu/abs/2011ASSP...21..229H}
}

@article{Bonetti:2018tpf,
    author = "Bonetti, Matteo and Sesana, Alberto and Haardt, Francesco and Barausse, Enrico and Colpi, Monica",
    title = "{Post-Newtonian evolution of massive black hole triplets in galactic nuclei \textendash{} IV. Implications for LISA}",
    eprint = "1812.01011",
    archivePrefix = "arXiv",
    primaryClass = "astro-ph.GA",
    doi = "10.1093/mnras/stz903",
    journal = "Mon. Not. Roy. Astron. Soc.",
    volume = "486",
    number = "3",
    pages = "4044--4060",
    year = "2019"
}

@ARTICLE{2019MNRAS.486.2336D,
       author = {{Dayal}, Pratika and {Rossi}, Elena M. and {Shiralilou}, Banafsheh and {Piana}, Olmo and {Choudhury}, Tirthankar Roy and {Volonteri}, Marta},
        title = "{The hierarchical assembly of galaxies and black holes in the first billion years: predictions for the era of gravitational wave astronomy}",
      journal = {\mnras},
         year = 2019,
        month = jun,
       volume = {486},
       number = {2},
        pages = {2336-2350},
          doi = {10.1093/mnras/stz897},
archivePrefix = {arXiv},
       eprint = {1810.11033},
 primaryClass = {astro-ph.GA},
       adsurl = {https://ui.adsabs.harvard.edu/abs/2019MNRAS.486.2336D}
}

@article{Salcido:2016oor,
    author = "Salcido, Jaime and Bower, Richard G. and Theuns, Tom and McAlpine, Stuart and Schaller, Matthieu and Crain, Robert A. and Schaye, Joop and Regan, John",
    title = "{Music from the heavens \textendash{} gravitational waves from supermassive black hole mergers in the EAGLE simulations}",
    eprint = "1601.06156",
    archivePrefix = "arXiv",
    primaryClass = "astro-ph.GA",
    doi = "10.1093/mnras/stw2048",
    journal = "Mon. Not. Roy. Astron. Soc.",
    volume = "463",
    number = "1",
    pages = "870--885",
    year = "2016"
}

@article{EPTA:2015ike,
    author = "Caballero, R. N. and others",
    collaboration = "EPTA",
    title = "{The noise properties of 42 millisecond pulsars from the European Pulsar Timing Array and their impact on gravitational wave searches}",
    eprint = "1510.09194",
    archivePrefix = "arXiv",
    primaryClass = "astro-ph.IM",
    doi = "10.1093/mnras/stw179",
    journal = "Mon. Not. Roy. Astron. Soc.",
    volume = "457",
    number = "4",
    pages = "4421--4440",
    year = "2016"
}

@article{Cornish:2017vip,
    author = "Cornish, Neil and Robson, Travis",
    editor = "Giardini, Domencio and Jetzer, Philippe",
    title = "{Galactic binary science with the new LISA design}",
    eprint = "1703.09858",
    archivePrefix = "arXiv",
    primaryClass = "astro-ph.IM",
    doi = "10.1088/1742-6596/840/1/012024",
    journal = "J. Phys. Conf. Ser.",
    volume = "840",
    number = "1",
    pages = "012024",
    year = "2017"
}

@article{Jenet:2006sv,
    author = "Jenet, F. A. and Hobbs, G. B. and van Straten, W. and Manchester, R. N. and Bailes, M. and Verbiest, J. P. W. and Edwards, R. T. and Hotan, A. W. and Sarkissian, J. M. and Ord, S. M.",
    title = "{Upper bounds on the low-frequency stochastic gravitational wave background from pulsar timing observations: Current limits and future prospects}",
    eprint = "astro-ph/0609013",
    archivePrefix = "arXiv",
    doi = "10.1086/508702",
    journal = "Astrophys. J.",
    volume = "653",
    pages = "1571--1576",
    year = "2006"
}

@article{Cornish:2001bb,
    author = "Cornish, Neil J.",
    title = "{Detecting a stochastic gravitational wave background with the Laser Interferometer Space Antenna}",
    eprint = "gr-qc/0106058",
    archivePrefix = "arXiv",
    doi = "10.1103/PhysRevD.65.022004",
    journal = "Phys. Rev. D",
    volume = "65",
    pages = "022004",
    year = "2002"
}

@article{Agarwal:2025cag,
    author = "Agarwal, Nikita and others",
    title = "{The NANOGrav 15 yr Data Set: Targeted Searches for Supermassive Black Hole Binaries}",
    eprint = "2508.16534",
    archivePrefix = "arXiv",
    primaryClass = "astro-ph.HE",
    month = "8",
    year = "2025"
}

@article{Xin:2021mmk,
    author = "Xin, Chengcheng and Haiman, Zoltan",
    title = "{Ultra-short-period massive black hole binary candidates in LSST as LISA {\textquoteleft}verification binaries{\textquoteright}}",
    eprint = "2105.00005",
    archivePrefix = "arXiv",
    primaryClass = "astro-ph.HE",
    doi = "10.1093/mnras/stab1856",
    journal = "Mon. Not. Roy. Astron. Soc.",
    volume = "506",
    number = "2",
    pages = "2408--2417",
    year = "2021"
}

@article{Xin:2024fci,
    author = "Xin, Chengcheng and Haiman, Zoltan",
    title = "{Identifying the electromagnetic counterparts of LISA massive black hole binaries in archival LSST data}",
    eprint = "2403.18751",
    archivePrefix = "arXiv",
    primaryClass = "astro-ph.HE",
    doi = "10.1093/mnras/stae2009",
    journal = "Mon. Not. Roy. Astron. Soc.",
    volume = "533",
    number = "3",
    pages = "3164--3173",
    year = "2024"
}

@article{Xin:2025voy,
    author = "Xin, Chengcheng and Isi, Maximiliano and Farr, Will M. and Haiman, Zolt{\'a}n",
    title = "{Identifying Compact Chirping SMBHBs in LSST using Bayesian Analysis}",
    eprint = "2506.10846",
    archivePrefix = "arXiv",
    primaryClass = "astro-ph.HE",
    month = "6",
    year = "2025"
}

@article{Burke-Spolaor:2018ghm,
    author = "Burke-Spolaor, Sarah and Blecha, Laura and Bogdanovic, Tamara and Comerford, Julia M. and Lazio, T. Joseph W. and Liu, Xin and Maccarone, Thomas J. and Pesce, Dominic and Shen, Yue and Taylor, Greg",
    title = "{The Next-Generation Very Large Array: Supermassive Black Hole Pairs and Binaries}",
    eprint = "1808.04368",
    archivePrefix = "arXiv",
    primaryClass = "astro-ph.GA",
    month = "8",
    year = "2018"
}

@ARTICLE{2024A&A...686A.183I,
       author = {{Izquierdo-Villalba}, David and {Sesana}, Alberto and {Colpi}, Monica and {Spinoso}, Daniele and {Bonetti}, Matteo and {Bonoli}, Silvia and {Valiante}, Rosa},
        title = "{Connecting low-redshift LISA massive black hole mergers to the nHz stochastic gravitational wave background}",
      journal = {\aap},
         year = 2024,
        month = jun,
       volume = {686},
          eid = {A183},
        pages = {A183},
          doi = {10.1051/0004-6361/202449293},
archivePrefix = {arXiv},
       eprint = {2401.10983},
 primaryClass = {astro-ph.GA},
       adsurl = {https://ui.adsabs.harvard.edu/abs/2024A&A...686A.183I}
}

@article{Antoniadis:2022pcn,
    author = "Antoniadis, J. and others",
    title = "{The International Pulsar Timing Array second data release: Search for an isotropic gravitational wave background}",
    eprint = "2201.03980",
    archivePrefix = "arXiv",
    primaryClass = "astro-ph.HE",
    doi = "10.1093/mnras/stab3418",
    journal = "Mon. Not. Roy. Astron. Soc.",
    volume = "510",
    number = "4",
    pages = "4873--4887",
    year = "2022"
}

@ARTICLE{2020ARA&A..58..257G,
       author = {{Greene}, Jenny E. and {Strader}, Jay and {Ho}, Luis C.},
        title = "{Intermediate-Mass Black Holes}",
      journal = {\araa},
         year = 2020,
        month = aug,
       volume = {58},
        pages = {257-312},
          doi = {10.1146/annurev-astro-032620-021835},
archivePrefix = {arXiv},
       eprint = {1911.09678},
 primaryClass = {astro-ph.GA},
       adsurl = {https://ui.adsabs.harvard.edu/abs/2020ARA&A..58..257G}
}

@ARTICLE{2012AdAst2012E..15N,
       author = {{Neumayer}, Nadine and {Walcher}, C. Jakob},
        title = "{Are Nuclear Star Clusters the Precursors of Massive Black Holes?}",
      journal = {Advances in Astronomy},
         year = 2012,
        month = jan,
       volume = {2012},
          eid = {709038},
        pages = {709038},
          doi = {10.1155/2012/709038},
archivePrefix = {arXiv},
       eprint = {1201.4950},
 primaryClass = {astro-ph.CO},
       adsurl = {https://ui.adsabs.harvard.edu/abs/2012AdAst2012E..15N}
}

@article{Kaaret:2017tcn,
    author = "Kaaret, Philip and Feng, Hua and Roberts, Timothy P.",
    title = "{Ultraluminous X-Ray Sources}",
    eprint = "1703.10728",
    archivePrefix = "arXiv",
    primaryClass = "astro-ph.HE",
    doi = "10.1146/annurev-astro-091916-055259",
    journal = "Ann. Rev. Astron. Astrophys.",
    volume = "55",
    pages = "303--341",
    year = "2017"
}

@article{Melchor:2025fey,
    author = "Melchor, Denyz and Naoz, Smadar and Gezari, Suvi and Mockler, Brenna",
    title = "{Tidal Disruption Event Demographics in Supermassive Black Hole Binaries over Cosmic Times}",
    eprint = "2507.08082",
    archivePrefix = "arXiv",
    primaryClass = "astro-ph.GA",
    doi = "10.3847/2041-8213/ae0a2e",
    journal = "Astrophys. J. Lett.",
    volume = "992",
    number = "2",
    pages = "L21",
    year = "2025"
}

@article{Barrows:2019pxt,
    author = "Barrows, R. Scott and Mezcua, Mar and Comerford, Julia M.",
    title = "{A Catalog of Hyper-luminous X-ray Sources and Intermediate-Mass Black Hole Candidates out to High Redshifts}",
    eprint = "1907.08213",
    archivePrefix = "arXiv",
    primaryClass = "astro-ph.GA",
    doi = "10.3847/1538-4357/ab338a",
    journal = "Astrophys. J.",
    month = "7",
    year = "2019"
}

@misc{enterprise,
  author       = {Justin A. Ellis and Michele Vallisneri and Stephen R. Taylor and Paul T. Baker},
  title        = {ENTERPRISE: Enhanced Numerical Toolbox Enabling a Robust PulsaR Inference SuitE},
  month        = sep,
  year         = 2020,
  howpublished = {Zenodo},
  doi          = {10.5281/zenodo.4059815},
  url          = {https://doi.org/10.5281/zenodo.4059815}
}

@article{EPTA:2023gyr,
    author = "Antoniadis, J. and others",
    collaboration = "EPTA, InPTA",
    title = "{The second data release from the European Pulsar Timing Array - V. Search for continuous gravitational wave signals}",
    eprint = "2306.16226",
    archivePrefix = "arXiv",
    primaryClass = "astro-ph.HE",
    doi = "10.1051/0004-6361/202348568",
    journal = "Astron. Astrophys.",
    volume = "690",
    pages = "A118",
    year = "2024"
}

@article{Zhao:2025pgg,
    author = "Zhao, Shi-Yi and others",
    title = "{Searching for Continuous Gravitational Waves in the Parkes Pulsar Timing Array Data Release 3}",
    eprint = "2508.13944",
    archivePrefix = "arXiv",
    primaryClass = "gr-qc",
    doi = "10.3847/1538-4357/ae0719",
    journal = "Astrophys. J.",
    volume = "992",
    number = "2",
    pages = "181",
    year = "2025"
}

@article{Baibhav:2020tma,
    author = "Baibhav, Vishal and Berti, Emanuele and Cardoso, Vitor",
    title = "{LISA parameter estimation and source localization with higher harmonics of the ringdown}",
    eprint = "2001.10011",
    archivePrefix = "arXiv",
    primaryClass = "gr-qc",
    doi = "10.1103/PhysRevD.101.084053",
    journal = "Phys. Rev. D",
    volume = "101",
    number = "8",
    pages = "084053",
    year = "2020"
}

@ARTICLE{2025ApJ...986..165T,
       author = {{Taylor}, Anthony J. and {Finkelstein}, Steven L. and {Kocevski}, Dale D. and {Jeon}, Junehyoung and {Bromm}, Volker and {Amor{\'\i}n}, Ricardo O. and {Arrabal Haro}, Pablo and {Backhaus}, Bren E. and {Bagley}, Micaela B. and {Banados}, Eduardo and {Bhatawdekar}, Rachana and {Brooks}, Madisyn and {Calabr{\`o}}, Antonello and {Ch{\'a}vez Ortiz}, {\'O}scar A. and {Cheng}, Yingjie and {Cleri}, Nikko J. and {Cole}, Justin W. and {Davis}, Kelcey and {Dickinson}, Mark and {Donnan}, Callum and {Dunlop}, James S. and {Ellis}, Richard S. and {Fern{\'a}ndez}, Vital and {Fontana}, Adriano and {Fujimoto}, Seiji and {Giavalisco}, Mauro and {Grazian}, Andrea and {Guo}, Jingsong and {Hathi}, Nimish P. and {Holwerda}, Benne W. and {Hirschmann}, Michaela and {Inayoshi}, Kohei and {Kartaltepe}, Jeyhan S. and {Khusanova}, Yana and {Koekemoer}, Anton M. and {Kokorev}, Vasily and {Larson}, Rebecca L. and {Leung}, Gene C.~K. and {Lucas}, Ray A. and {McLeod}, Derek J. and {Napolitano}, Lorenzo and {Onoue}, Masafusa and {Pacucci}, Fabio and {Papovich}, Casey and {P{\'e}rez-Gonz{\'a}lez}, Pablo G. and {Pirzkal}, Nor and {Somerville}, Rachel S. and {Trump}, Jonathan R. and {Wilkins}, Stephen M. and {Yung}, L.~Y. Aaron and {Zhang}, Haowen},
        title = "{Broad-line AGNs at 3.5 < z < 6: The Black Hole Mass Function and a Connection with Little Red Dots}",
      journal = {\apj},
         year = 2025,
        month = jun,
       volume = {986},
       number = {2},
          eid = {165},
        pages = {165},
          doi = {10.3847/1538-4357/add15b},
archivePrefix = {arXiv},
       eprint = {2409.06772},
 primaryClass = {astro-ph.GA},
       adsurl = {https://ui.adsabs.harvard.edu/abs/2025ApJ...986..165T}
}

@article{Matthee:2023utn,
    author = "Matthee, Jorryt and others",
    title = "{Little Red Dots: An Abundant Population of Faint Active Galactic Nuclei at z {\ensuremath{\sim}} 5 Revealed by the EIGER and FRESCO JWST Surveys}",
    eprint = "2306.05448",
    archivePrefix = "arXiv",
    primaryClass = "astro-ph.GA",
    doi = "10.3847/1538-4357/ad2345",
    journal = "Astrophys. J.",
    volume = "963",
    number = "2",
    pages = "129",
    year = "2024"
}

@article{Kato:2023tfz,
    author = "Kato, Ryo and Takahashi, Keitaro",
    title = "{Precision of localization of single gravitational-wave source with pulsar timing array}",
    eprint = "2308.10419",
    archivePrefix = "arXiv",
    primaryClass = "gr-qc",
    doi = "10.1103/PhysRevD.108.123535",
    journal = "Phys. Rev. D",
    volume = "108",
    number = "12",
    pages = "123535",
    year = "2023"
}

@article{Kato:2025set,
    author = "Kato, Ryo and Takahashi, Keitaro",
    title = "{Realistic assessment of a single gravitational wave source localization taking into account precise pulsar distances with pulsar timing arrays}",
    eprint = "2506.02819",
    archivePrefix = "arXiv",
    primaryClass = "gr-qc",
    month = "6",
    year = "2025"
}

@article{Furusawa:2025yuc,
    author = "Furusawa, Kazuya and Kuroyanagi, Sachiko and Ichiki, Kiyotomo",
    title = "{Resolving Individual Signals in the Presence of Stochastic Background in Future Pulsar Timing Arrays}",
    eprint = "2505.10284",
    archivePrefix = "arXiv",
    primaryClass = "astro-ph.CO",
    doi = "10.1093/mnras/staf1497",
    journal = "Mon. Not. Roy. Astron. Soc.",
    volume = "1010",
    pages = "1022",
    year = "2025"
}
\end{document}